\documentclass[twocolumn]{aastex63}
\usepackage{float}

\newcommand\xmm{{\it XMM-Newton}}
\newcommand\swift{{\it Swift}}

\newcommand\s{{\rm~s}}

\newcommand\kev{{\rm~keV}}

\newcommand\kms{\ifmmode {\rm~km\ s}$^{-1}$ \else ~km s$^{-1}$\fi}
\newcommand\Hunit{\ifmmode {\rm~km\ s}$^{-1}$\ {\rm Mpc}$^{-1}$
        \else ~km s$^{-1}$ Mpc$^{-1}$\fi}
\newcommand\ctssec{\ifmmode {\rm~count\ s}$^{-1}$ \else ~count s$^{-1}$\fi}
\newcommand\ergsec{\ifmmode {\rm~erg\ s}$^{-1}$ \else
        ~erg s$^{-1}$\fi}
\newcommand\funit{\ifmmode {\rm~erg\ s}$^{-1}$\ ; {\rm cm}$^{-2}$ \else
        ~ergs s$^{-1}$ cm$^{-2}$\fi}
\newcommand\phflux{\ifmmode {\rm~photon\ s}$^{-1}$\  ; {\rm cm}$^{-2}$
        \else   ~photon s$^{-1}$ cm$^{-2}$\fi}
\newcommand\efluxA{\ifmmode {\rm~erg\ s}$^{-1}$\ ; {\rm cm}$^{-2}$\ ; {\rm
        \AA}$^{-1}$ \else ~erg s$^{-1}$ cm$^{-2}$ \AA$^{-1}$\fi}
\newcommand\efluxHz{\ifmmode {\rm~erg\ s}$^{-1}$\ ; {\rm cm}$^{-2}$\ ; {\rm
        Hz}$^{-1}$ \else ~erg s$^{-1}$ cm$^{-2}$ Hz$^{-1}$\fi}
\newcommand\cc{\ifmmode {\rm~cm}$^{-3}$ \else cm$^{-3}$\fi}
\newcommand\FWHM{\ifmmode {\rm~FWHM} \else ${\rm~FWHM}$\fi}
\newcommand\Msun{\ifmmode M_{\odot} \else $M_{\odot}$\fi}
\newcommand\Lsun{\ifmmode L_{\odot} \else $L_{\odot}$\fi}

\newcommand\hbeta{\ifmmode {\rm H}\beta \else H$\beta$\fi}
\newcommand\Kalpha{\ifmmode {\rm K}\alpha \else K$\alpha$\fi}
\newcommand\nh{\ifmmode N_{\rm H} \else N$_{\rm H}$\fi}
\makeatletter

\newcommand{\Rmnum}[1]{\expandafter\@slowromancap\romannumeral #1@}
\makeatother

\def\kev{\mbox{keV}}

\def\Msun{\mbox{M$_\odot$}}

\def\kms{{\rm\,km\,s^{-1}}}

\def\mathnew{\mathsurround=0pt}   
\def\simov#1#2{\lower .5pt\vbox{\baselineskip0pt  
\lineskip-.5pt\ialign{$\mathnew#1\hfil##\hfil$\crcr#2\crcr\sim\crcr}}}

\def\'#1{\ifx#1\i{\accent"13\i}\else{\accent"13#1}\fi}    
\def\eg{e.g.,}  
     
\def\ie{i.e.} 
\def\b{\textcolor{blue}}
\def\r{\textcolor{red}}

\begin{document}

\shorttitle{Strong Soft excess in OJ~287} \shortauthors{Pal, Main et al. 2019}

\title{Strong Soft X-ray Excess in 2015 \emph{XMM-Newton} Observation
of BL--Lac OJ~287}

\correspondingauthor{M.~Pal}
\email{rajanmainpal@gmail.com,pankaj.tifr@gmail.com}

\author[0000-0001-6523-6522]{Main Pal}
\affiliation{Centre for Theoretical Physics, Jamia Millia Islamia, New Delhi 110025, India}

\author[0000-0002-0786-7307]{Pankaj Kushwaha}
\affiliation{Department of Astronomy (IAG-USP), University of Sao Paulo, Sao Paulo 05508-090, Brazil}
\affiliation{Aryabhatta Research Institute of Observational Sciences (ARIES), Nainital 263002, India }
\altaffiliation{Aryabhatta Postdoctoral Fellow}

\author[0000-0003-1589-2075]{G.C. Dewangan}
\affiliation{Inter University Centre for Astronomy and Astrophysics (IUCAA), Pune 411007, India}


\author{P. K. Pawar}
\affiliation{Inter University Centre for Astronomy and Astrophysics (IUCAA), Pune 411007, India}

\begin{abstract}
{We report a strong soft X-ray excess in the BL--Lacartae $\gamma$-ray
blazar OJ~287 during the long exposure in May 2015, amongst two of the
latest \xmm~{} observations performed in May 2015 and 2018. In case
of May 2015 observation, a logparabola model fits the EPIC-pn data well
while a logparabola plus powerlaw describes the overall simultaneous optical
to X-ray spectra, suggesting the excess as the synchrotron tail. This interpretation,
however, is inconsistent with the observed spectral break between near-infrared and
optical spectra, attributed to standard disk around a
supermassive black hole (SMBH). Based on this, we considered two commonly invoked accretion
disk based models in AGNs to explain the soft excess: the cool Comptonization component
in the accretion disk and the blurred reflection from the partially ionized accretion
disk. We found that both cool Comptonization and blurred reflection models provide
equally good fit to the data and favor a super-heavy SMBH of mass $\sim 10^{10}~
M_\odot$. Further investigation of about a month long simultaneous
X-ray and UV pointing observations revealed a delayed UV emission with respect to
the 1.5-10 keV band, favoring X-ray reprocessing phenomenon as the dominant mechanism.
The
results suggest that the soft excess is probably caused by strong light bending
close to the SMBH. The detected soft excess in 2015 data and
its disappearance in 2018 data is also consistent with the presence of accretion
disk emission, inferred from the NIR-optical spectral break between May 2013 to March 2016.
}
\end{abstract}

\keywords{galaxies:  active --- galaxies: jets-BL Lacs ---indivdual: OJ~287 ---
   radiation mechanisms: non-thermal-X-rays}
\section{Introduction} \label{sec:intro}
OJ~287 $(z=0.306)$ is one of the most luminous and rapidly variable BL-Lacartae
objects (BLLs) at radio to optical frequencies \citep{1985PASP...97.1158S,1989A&AS...80..103S}.
It is also one of the most extensively studied extra-galactic active galactic nuclei
(AGNs) over the entire electromagnetic spectrum from radio to $\gamma$-rays \citep[and
references therein]{1973ApJ...179..721V,2013A&A...559A..20H,2016ApJ...819L..37V,
2017MNRAS.465.4423G,2018MNRAS.478.3199B,2018ApJ...863..175G,2018MNRAS.473.1145K,
2018MNRAS.479.1672K,2018MNRAS.480..407K}.
Apart from the typical stochastic variability of blazars and favorable observational
properties like high radio and optical brightness, the most prominent features
responsible for making the source famous is the presence of a recurrent regular
optical outbursts every $\sim 12$-yr \citep{1988ApJ...325..628S,1996A&A...305L..17S}
and its double-peaked structure \citep{1996A&A...315L..13S}. 

Two interpretations have been suggested in the literature for the regular optical outbursts.
One class of models attribute the quasi-periodic outbursts to the interaction dynamics of accretion
disk and SMBHs \citep{1988ApJ...325..628S,1996ApJ...460..207L} in a binary SMBH
system while the other class of models attributes it to the Doppler boosted jet
emission as a consequence of geometrical alignment of precessing single \citep[and references
therein]{,2018MNRAS.478.3199B} or double relativistic jets \citep[and references therein]
{2018arXiv181111514Q}. The very first model by \citet{1988ApJ...325..628S} explained
the periodicity to increased accretion flow due to tidal disturbances induced
by the secondary SMBH in the accretion disk of primary SMBH. The model was modified
after the observation of sharp rise during the 1994 and 1996 outbursts by \citet{1996ApJ...460..207L}
who attributed the periodicity and double-peaked structure as the impact of secondary SMBH
on the primary accretion disk. The disk-impact binary SMBH model has been fairly
successful in predicting the timing of the double-peaked $\sim$ 12-yr quasi-periodic
outbursts \citep{2016ApJ...819L..37V, 2013A&A...559A..20H}. It attributes the flare
emission to thermal bremsstrahlung of the hot gas torn off during the impact and constrains
the SMBH masses to $\sim 1.8\times10^{10}~M_{\odot}$ and $\sim 1.5\times10^8~M_{\odot}$
for the primary and secondary SMBHs, respectively \citep{2012MNRAS.427...77V,2016ApJ...819L..37V}.
The geometrical class of models, on the other hand, argue a total system mass in the range
of a few times $\sim 10^7- 10^9~M_{\odot}$ \citep[and references therein]{2007Ap&SS.310...59S,2018MNRAS.478.3199B,2018arXiv181111514Q}.

From the shape of the broadband energy spectra, it is known that that OJ 287 is a low-peaked
BLL with the peak of the low-energy hump, attributed to synchrotron emission from
the jet, at near-infrared (NIR) energies. The high-energy hump in the X-ray to 
$\gamma$-ray band normally peaks at $\sim 100$ MeV \citep{2010ApJ...716...30A,2013MNRAS.433.2380K}.
 The synchrotron-self Compton (SSC) mechanism successfully describes it
typical X-ray emission while $\gamma$-ray emission is shown to be due to the inverse
Comptonization of a $\sim$250 K {($\sim 0.022$ eV)} torus photon field \citep[EC-IR][]{2013MNRAS.433.2380K},
contrary to the generally believed SSC origin of high-energy hump in BLLs. However,
during its latest multi-wavelength activity from December 2015 - 2017, OJ~287 exhibited
a hardened MeV-GeV emission, showing a clear shift in the peak of the high-energy hump
to GeV energies \citep{2018MNRAS.473.1145K,2018MNRAS.479.1672K}. 
At the same time,
a spectral break between NIR-optical emission was also observed for the first-time
as reported by \citet{2018MNRAS.473.1145K}. The occurrence of NIR-optical
spectral break was traced back to May 2013 (MJD 57439) and continued since then
till March 2016. They further showed that the observed MeV-GeV spectral change
can be naturally reproduced
by external Comptonization but this time by IC of broad line region photons
\citep{2018MNRAS.473.1145K} which have been detected during the previous cycles
of $\sim$12-yr optical outbursts \citep{2010A&A...516A..60N}. The NIR-optical spectral break is most naturally explained by the sandard disk emission of a $\sim 10^{10}~M_\odot$ SMBH. Interestingly, its first appearance in May
2013 \citep[MJD 56439;][]{2018MNRAS.473.1145K,2019BHCB} is very close to
the impact time predicted in the disk-impact binary SMBH model \citep{1996ApJ...460..207L}
in the BH frame. This spectral and temporal coincidence currently tilts
the central engine debate in the favor of disk-impact binary SMBH model.

 Survey of literature records show that OJ 287 has exhibited the most dramatic
spectral variations in X-ray energy band. The reported spectral shapes cover all
the possible energy-spectral profiles, from a powerlaw -- the typical X-ray
spectrum of OJ 287 \citep[]{2009PASJ...61.1011S,2010ApJ...716...30A,2013MNRAS.433.2380K}
to flat
ones, \citep[e.g.][]{2017MNRAS.468..426S,2018MNRAS.479.1672K}, extremely soft spectra
\citep[e.g.][]{2001PASJ...53...79I,2018MNRAS.479.1672K} as well as a mixture of
these \citep[e.g.][]{2001PASJ...53...79I,2012MNRAS.427...77V,2018MNRAS.473.1145K}.
As already mentioned, the typical powerlaw X-ray energy-spectra are successfully described by SSC emission \citep{2009PASJ...61.1011S,
2013MNRAS.433.2380K} while the flat and mix spectra (typical+soft) have been argued
to be as a result of mixture of synchrotron and the SSC emission \citep{2001PASJ...53...79I,
2017MNRAS.468..426S} in one interpretation. The other possibility, argued but not
yet studied, is an additional spectral component like Bethe-Heitler emission. The
extremely soft X-ray spectra observed during the 2016-2017 activity is shown to be
a new, additional high-frequency-peaked BLL (HBL) emission component
by \citet{2018MNRAS.479.1672K}, thanks to the coordinated MW follow-ups. In light
of this, the extremely soft X-ray spectra \citep{2001PASJ...53...79I} which have
been observed earlier as well could be the HBL component. Interestingly, within the
limit of available records, strongly soft X-ray spectra seem to be a common feature
of the source, present for a few years around the $\sim 12$-yr quasi-periodic
optical outbursts. 

In this work, we perform spectral study of the 2015 and 2018 \xmm{}~observations
of OJ 287, supplementing with multiple \swift{}~XRT/UVOT observations to explore the soft
X-ray excess in the 2015 \xmm{}~data. In the next section, we present
details of observation and data reduction. \S3 presents the systematic
spectral analysis of data and results. In section 4, we report our discussion and
conclusions. We used the cosmological parameters $H_{0} = 67.04~{\rm km~s^{-1}~Mpc^{-1}}$,
$\Omega_m = 0.3183$ and $\Omega_{\Lambda} = 0.6817$ \footnote{http://www.kempner.net/cosmic.php}
to calculate the distance.

\section{Observation and data reduction}
OJ~287 has been observed multiple times by \xmm{}~observatory, mainly around
the {$\sim$ 12-yr} quasi-periodic optical outbursts. Some of the previous
observations have been studied
in detail by various authors \citep{2018MNRAS.473.3638G, 2016MNRAS.462.1508G}. The
latest observation with $\sim$28 ks and the longest exposure ($\sim 129$~ks) of this
object were performed in May 2018 (MJD=58149-58150) and May 2015 (MJD=57149-57150),
respectively. The European Photon Imaging Camera (EPIC)-pn \citep{turner2001}
was operated in the prime large window mode with the thin filter during both the observations.
We also used \swift{}~XRT/UVOT observations from MJD=57140.4 to MJD=57173.6. This period includes the
May 2015 \xmm{}~observation. 

We followed the standard reduction procedure using the \xmm{} Science Analysis System
({\tt SAS v15.0}) \citep{2004ASPC..314..759G} with latest calibration files. First, we
reprocessed the EPIC-pn data using {\tt epproc} and obtained event files. We removed
the intervals affected by flaring particle background by examining light curves
above $10\kev$ to get events file. We used single and double events (PATTERN $\le4$)
for the EPIC-pn, and omitted events at the CCD edges and bad pixels (FLAG=0). We extracted the
source spectrum using a circular region of 50 arcsec, centered at the
source. We also obtained a background spectrum from a circular region of the same size
away from the source and free from any sources. The resulted net exposures
were found to be $\sim$ 53 ks and $\sim$ 19 ks for the 2015 and 2018 observations, 
respectively. The net count rates in 2-10 keV band were observed to be $0.316\pm0.003$ and $0.299\pm0.004$
counts $s^{-1}$ for the 2015 and 2018 data sets, respectively.  We also examined the pileup very carefully using {\tt epatplot}. We
did not find any significant pileup which might affect our analysis. Finally, we generated
response matrix and ancillary response files at the source position using the tools
{\tt  rmfgen} and {\tt arfgen}, respectively. We grouped the data using the SAS task
{\tt specgroup} with an oversampling of $3$ and minimum counts of $20$ per bin.

\begin{figure}
\includegraphics[scale=0.33,angle=-90.0]{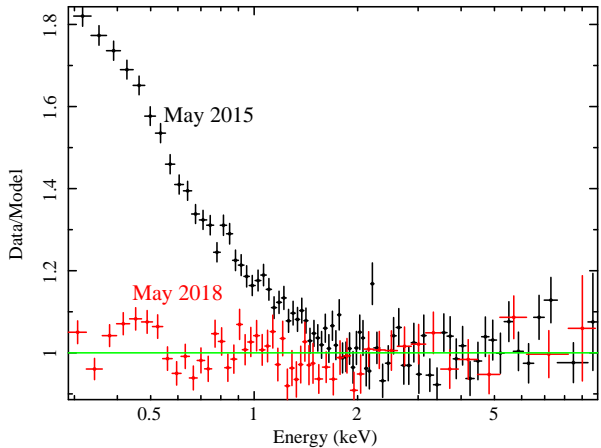}
\caption{The ratio (Data/Model) plot of~absorbed powerlaw model fitted in $2-10$ \kev
~band and then extended to $0.3-2$\kev. A clear soft excess is present at low energies
in the May 2015 while 2018 data seems consistent with a powerlaw model.}
  \label{fig1}	
\end{figure}

 For the reduction of \swift{}~XRT and UVOT data, we followed the steps described
in \citet{2018MNRAS.474.5351P}. We selected background annular region from 10 arcsec
to 20 arcsec centered at the source coordinates. We also omitted the data points from
bad patches of the CCD in case of UVOT observations.

\section{Data analysis}\label{dataAn}

\subsection{Spectral analysis}\label{subsec:spec}
\subsubsection{X-ray ray emission}
We used {\tt XSPEC v12.10.1} \citep{1996ASPC..101...17A} to analyze the X-ray spectra
of OJ~287 and used the $\chi^2$ statistics for the model fitting. Unless stated
otherwise, the errors on the best-fit parameters are quoted at 90\%~confidence level, corressponding to $\Delta
\chi^{2}$=2.706.

\begin{figure*}
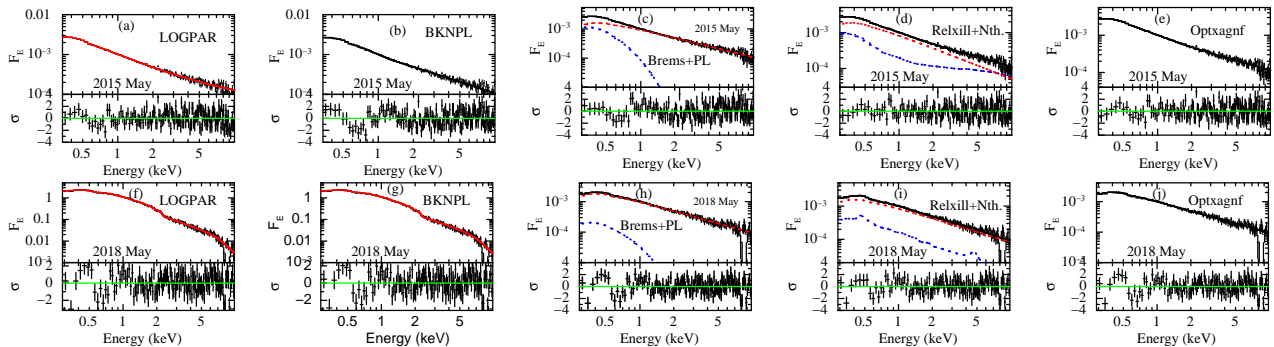
  
  \includegraphics[scale=0.268,angle=-90.0]{fig2a.eps}
  \includegraphics[scale=0.268,angle=-90.0]{fig2b.eps}
  \includegraphics[scale=0.268,angle=-90.0]{fig2c.eps}
  \includegraphics[scale=0.268,angle=-90.0]{fig2d.eps}
  \includegraphics[scale=0.268,angle=-90.0]{fig2e.eps}

  \includegraphics[scale=0.268,angle=-90.0]{fig2f.eps}
  \includegraphics[scale=0.268,angle=-90.0]{fig2g.eps}
  \includegraphics[scale=0.268,angle=-90.0]{fig2h.eps}
  \includegraphics[scale=0.268,angle=-90.0]{fig2i.eps}
  \includegraphics[scale=0.268,angle=-90.0]{fig2j.eps}

\caption{ The best--fit model, data in top panel and residuals (in $\sigma$) in bottom panel are shown for May 2015 (first row) and May 2018 (second row) for models {\tt tbabs$\times$(logpar)} (a,f), {\tt tbabs$\times$(bknpower)} (b,g), {\tt tbabs$\times$(bremss+powerlaw)} (c,h), {\tt tbabs$\times$(relxill+nthcomp)} (d, i) and {\tt tbabs$\times$(optxagnf)} (e, j), respectively
(see Section \ref{dataAn}). Solid line on the data points represents the net composite model while dashed lines are different model components of the net composite model. For all the models, residuals are within 2-3$\sigma$.}
\label{fig2}	
\end{figure*}

We considered only EPIC-pn data due to its high signal to noise ratio compared
to the EPIC-MOS. We began by fitting the 2-10 keV band with an absorbed powerlaw
({\tt tbabs$\times$powerlaw}) model. We fixed the absorption column to the Galactic value of $N_{H}= 3.04\times10^{20}~\rm
cm^{-2}$ \citep{dickey1990}. This resulted in a $\chi^{2}/\nu$ of 157.8/116 and 93.2/97
for the May 2015 and 2018 data respectively, where $\nu$ stands for the degree of
freedom. The best-fit power-law photon index $\Gamma$ was found to be $1.91\pm0.03$
and $2.06\pm0.06$ for the two data sets. Thus, both the data sets represent
a different spectral state of the source. We then extrapolated the best-fit model down
to 0.3 keV for both the observations as shown in Fig.~\ref{fig1}. Surprisingly,
the May 2015 data revealed a strong soft X-ray excess, observed rarely in BLLs,
  but commonly observed in radio quite AGN such as the narrow-line Seyfert
type 1 (NLS1)\citep{1999ApJS..125..317L,crum2006,2004MNRAS.349L...7G}.
To investigate this, we then systematically fitted the whole range (0.3--10 keV)
with possible phenomenological models, starting first with log-parabola
({\tt logpar}) and broken power-law emission ({\tt bknpower}) and then additionally
thermal bremsstrahlung ({\tt bremss}) as per claims in the literature \citep{2016ApJ...819L..37V}
and finally the two AGNs models: cool Comptonization
\citep[:{\tt optxagn}]{2012MNRAS.420.1848D} and blurred reflection
\citep[: \ie~{\tt relxill}]{2013MNRAS.430.1694D,2004MNRAS.349.1435M}.

A simple {\tt powerlaw} model fit to the 2015 data over 0.3--10 keV band resulted
in a poor fit ($\chi^{2}/\nu=893.7/160$) due to the presence of the strong soft excess.
Since this AGN is a blazar, the X-ray emission may be described phenomenologically
by {\tt logpar} and {\tt bknpower}, independently. Fitting {\tt tbabs$\times$logpar}
over 0.3--10 keV band resulted in $\chi^{2}/\nu=202.8/159$ while {\tt tbabs$\times$bknpower}
fit resulted in $\chi^{2}/\nu=234.6/158$. The best-fit parameters for both models
are listed in Table~\ref{table1} and the corresponding data, model, and residuals
(in $\sigma$) are shown in Fig.~\ref{fig2}~(a) and \ref{fig2}~(b). The {\tt logpar}
fit describes the data fairly well, however, it is not consistent with the
broadband emission of the source (see \S\ref{discussion} and Fig. \ref{fig5}). 

Another claim is thermal bremsstrahlung radiation \citep{2016ApJ...819L..37V}
from a $3\times10^5$ K gas \citep{2012MNRAS.427...77V} around the expected $\sim$12-yr
quasi-periodic optical outbursts. This temperature corresponds to $\sim$ 25 eV and is
irrelevant for the observed soft X-ray excess. Nonetheless, we, additionally
explored a redshifted {\tt zbremss} model along with the above considered models.
This model has three parameters--plasma temperature, normalization, and
the source redshift. We allowed plasma temperature and its normalization to vary.
The fit resulted in $\chi^{2}/\nu=211.8/159$. The best-fitting parameters are listed
in Table~\ref{table1} while the plots are shown in top and bottom panels of Fig.
\ref{fig2}~(c). During model fitting, we found an statistically acceptable fit
with a 25 keV plasma temperature. However, this temperature is too
high, dominating the high energy end of the X-ray and is contrary to
the general behavior of blazars.

Since an accretion disk emission has been claimed for the NIR-optical spectral
break in a systematic analysis by \citet{2018MNRAS.479.1672K}, we invoked
disk-based soft excess models used to explain the soft X-ray excess normally seen in
Seyfert type 1 AGN \citep{crum2006,2004MNRAS.349L...7G,2012MNRAS.420.1848D}.
Though the origin is still unclear, two
competing models -- blurred reflection and cool Comptonization have been most
acceptable. Thus, to a simple absorbed {\tt powerlaw} model we added a reflection
model {\tt relxill} which is a combination of {\tt xillver} \citep{ 2011ApJ...731..131G,
2013ApJ...768..146G} and {\tt relline} \citep{2010MNRAS.409.1534D, 2013MNRAS.430.1694D}.
This model calculates the reflected emission at each angle at each radius of the
accretion disc \citep{2014ApJ...782...76G}. The details of parameters of {\tt relxill}
and its different application forms are described briefly on the webpage 
document\footnote{\url{http://www.sternwarte.uni-erlangen.de/~dauser/research/relxill/}}.

The applied form of {\tt relxill} assumes that the X-ray source illuminates the
accretion disc in a lamppost geometry \citep{2004MNRAS.349.1435M}. The illumination
is described as a broken emissivity law which has the form $\epsilon \propto
r^{-q_{in}}$ between $r_{in}$ and $r_{br}$; $\epsilon \propto r^{-q_{out}}$ between
$r_{br}$ and $r_{out}$; where $r$ is the radius of the accretion disk,
 $q_{in}$, and $q_{out}$ are inner and outer emissivity indices; $r_{in}$, $r_{br}$, and
$r_{out}$ are the inner, break and outer radii of the accretion disk. The other parameters
are spin ($a$), inclination angle ($i$), iron abundance $A_{Fe}$ relative to solar
abundance, illuminating power--law index ($\Gamma$), high energy cutoff ($E_{cut}$),
ionization parameter ($\xi=L/nr^{2}$ with $L$ being the source X-ray luminosity
and $n$ is the hydrogen number density of the disk material) and reflected fraction
denoted by $R$. We fixed the iron abundance to 1, the inclination
to 3$^\circ$, high energy cutoff to 300\kev~and outer radius to $400r_{g}$. We tied
the {\tt relxill} photon index $\Gamma$ to powerlaw photon index $\Gamma$ and hence
the $R$ parameter was fixed to $-1$ under the lamppost scenario. We allowed the rest
of the parameters and the fit with {\tt tbabs$\times$(relxill+powerlaw)} model resulted
in $\chi^{2}/\nu=184.6/154$. To be more realistic, we replaced the phenomenological
powerlaw model by {\tt nthcomp} \citep{1996MNRAS.283..193Z, 1999MNRAS.309..561Z}
which can correctly predict the low energy rollover where Galactic absorption can
modify the spectrum. We fixed the seed photon temperature at 2 eV and electron temperature
associated with X-ray corona to 100 \kev. The fit resulted in $\chi^{2}/\nu= 184.8/154$
with results listed in Table~\ref{table1} and corresponding plots in the top and bottom
panels of Fig.~\ref{fig2}~(d), respectively.

\begin{table}
\caption{Best-fitting parameters of models used to fit the 0.3-10~\kev~band for two \xmm{}~observations. The flux f$_{E}$ is measured in units of $10^{-12}$\funit.  ``t" stands for paramter tied to other parameter.}
\label{table1} 
\begin{tabular}{lccccc}
\hline 
\hline 
\small
Model component        &2015 May & 2018 May   \\
\hline
$N_{H(\rm Galaxy)}$ ($10^{20}\rm cm^{-2}$)  &3.04~(f)&3.04~(f)\\ 
\hline 
& \multicolumn{2}{c}{Logpar}\\

slope ($\alpha$) &$2.26\pm0.01$ &$2.09\pm0.01$    \\
curv. term ($\beta$)    &$-0.33\pm0.02$ &$-0.06\pm0.04$ \\                      
Norm.(LP) ($10^{-3}$)&$1.07\pm0.01$ &$1.04\pm0.01$  \\
Stat. ($\chi^{2}/\nu$) &202.8/159&168.9/138 \\
\hline 
& \multicolumn{2}{c}{Bknpower}\\
Photon index ($\Gamma_1$) &$2.38\pm0.03$&$2.11\pm0.02$ \\                               
Photon index ($\Gamma_2$) &$1.96\pm0.03$ &$2.02\pm0.05$ \\                              
$E_{break}$ (keV)&$1.3\pm0.1$&$1.67_{-0.4}^{+0.8}\pm0.03$  \\
Norm.(BPL)($10^{-3}$)&$1.06\pm0.02$ &$1.04\pm0.01$  \\ 
Stat. ($\chi^{2}/\nu$)&234.6/158&167.3/137 \\
\hline 

& \multicolumn{2}{c}{PL+bremss}\\
Photon index $\Gamma$ &$1.96\pm0.02$&$2.03_{-0.05}^{+0.04}$ \\
Norm. (nth) ($10^{-3}$)&$0.96\pm0.02$ &$1.00_{-0.06}^{+0.04}$ \\ 
plasma temp. ($\frac{\rm kT_{brem}}{\rm keV}$)  &$0.31\pm0.02$ &$0.42_{-0.28}^{+0.21}$\\
Norm. (brem) ($10^{-3}$)&$4.6_{-0.4}^{+0.5}$&$0.61_{-0.38}^{+0.15}$ \\ 
Stat. ($\chi^{2}/\nu$)&211.8/159&168.9/137\\
f$_{E}$~($0.3-2~\kev$) &$3.1$&$2.8$\\
f$_{E}$~($2-10~\kev$) &$2.64$&2.5\\

\hline 
& \multicolumn{2}{c}{Nth.+Relxill}\\
Photon index ($\Gamma$) &$2.21_{-0.12}^{+0.06}$ &$2.03_{-0.02}^{+0.04}$\\ 
Norm.(nth) ($10^{-3}$)&$0.86_{-0.05}^{+0.07}$ &$0.89_{-0.11}^{+0.12}$  \\
Index1   ($q_{in}$) &$7.1_{-2.2}^{+0.30}$           &$3$ (f)   \\
Index2 ($q_{out}$)    &$4.2_{-0.9}^{+0.1}$     &$3$ (f) \\
Photon index ($\Gamma$)&$2.21_{-0.12}^{+0.06}$ (t) &$2.03_{-0.02}^{+0.04}$ (t) \\
 log($\frac{\rm Ionization~par.}{\rm erg~cm~s^{-1}}$) &$2.3_{-0.5}^{+0.4}$ &$3.3_{-2.2}^{+0.3}$ \\

  Inner radius ($r_{g}$)&$1.6_{-0.1}^{+0.3}$ &$11.7_{-9.6}^{+7.0}$\\
  Break radius  ($r_{g}$) &$4.30_{-0.01}^{+1.6}$ &$11.7_{-9.6}^{+7.0}$ (t) \\
   Spin ($a$)  &$0.99_{-0.07}^{+0.01}$ &$0.998_{-0.954}^{+0.038}$ \\
Norm.(refl) ($10^{-5}$)&$2.6_{-1.6}^{+0.5}$ &$0.14\pm0.01$ \\
Stat. ($\chi^{2}/\nu$)&184.7/154&162.3/135\\
f$_{E}$~($0.3-2~\kev$)&$3.1$&$2.8$\\
f$_{E}$~($2-10~\kev$)&2.66&2.5 \\

\hline 
 & \multicolumn{2}{c}{optxagnf}\\
   Acc. rate ($\frac{\rm L}{\rm L_{edd}}$)   &$0.044_{-0.003}^{+0.016}$ &$0.0084_{-0.0011}^{+0.0038}$    \\ 
   Spin ($a$)    &$0.996^{+0.002}_{-0.095}$                                      &$0.99_{-0.25}^{+0.008}$    \\ 
   Coronal radius ($r_{g}$) &$6.1_{-1.3}^{+6.3}$                       &$6.9_{-0.3}^{+3.2}$    \\ 
  Plasma temp. ($\frac{\rm kT_e}{\rm keV})$  &$0.46_{-0.05}^{+0.17}$   &$0.37_{-0.18}^{+0.86}$    \\ 
   Optical depth ($\tau$)     &$9.0_{-1.6}^{+0.2}$                     &$12.1_{-9.0}^{+1.2}$    \\ 
   Frac. power ($f_{pl}$)     &$0.11_{-0.05}^{+0.01}$                  &$0.67_{-0.03}^{+0.29}$    \\ 
       Photon index ($\Gamma$)&$1.88_{-0.04}^{+0.05}$                  &$2.01_{-0.03}^{+0.07}$    \\
Stat. ($\chi^{2}/\nu$)&188.9/155                                       &168.1/134 \\
f$_{E}$~$(0.3-2~\kev)$&$3.1$&$2.8$\\
f$_{E}$~($2-10~\kev$) &2.66 &2.5\\

\hline  
\end{tabular}
\end{table}

\begin{figure*}
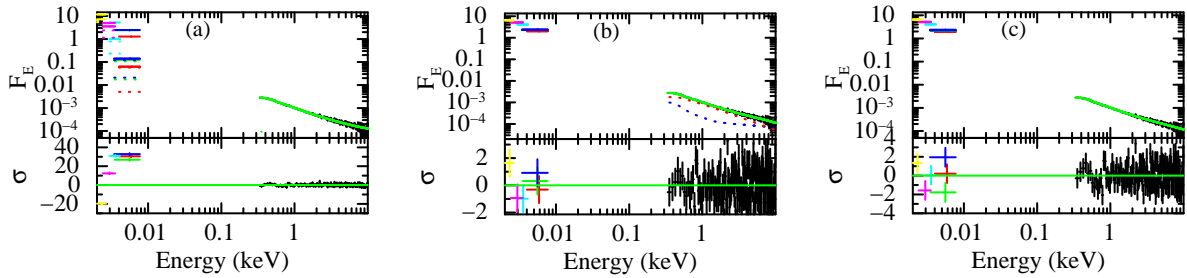

  \includegraphics[scale=0.422,angle=-90.0]{fig3a.eps}
  \includegraphics[scale=0.422,angle=-90.0]{fig3b.eps}
  \includegraphics[scale=0.422,angle=-90.0]{fig3c.eps}
\caption{ Best-fit model, data and residuals for (a) {\tt logpar+bremss}. The fit
shows strong positive residuals in UV/Optical bands (b) {\tt relxill+nthcomp}. (c)
{\tt optxagnf}. No significant residuals (within 2-3$\sigma$) is present in
UV/Optical bands in the best-fitting models related to reflection (b) and
Comptoniozation (c). Solid line on the data points represents the net composite
model used here and dashed lines are different model components of the net composite
model.}
  \label{fig3}	
\end{figure*}

The other widely argued scenario of AGNs soft X-ray excess attributes it to a
different plasma embedded in the interior region of the accretion disk. As argued
by \citet{2012MNRAS.420.1848D}
the gravitational potential energy is released at each point of the accretion disk
as a blackbody emission down to R$_{corona}$. Below this radius, the gravitational
potential energy is no longer completely thermalized. The energy is distributed into
two types of plasma-- an optically thick ($\tau>1$) cool plasma (kT$_e$$\sim$0.2 keV) for soft X-ray excess emission
and an optically thin ($\tau<1$) hot plasma (kT$_e$$\sim$ 100 keV) emitting power--law
continuum above 2 keV. Thus, to the {\tt powerlaw}, we added
{\tt optxagnf} model which incorporates the above mention scenario. The important
parameters of the model are -- accretion rate relative to the Eddington rate L/L$_{Edd}$,
mass of the BH M$_{BH}$ and its spin $a$, source luminosity distance D$_L$, cool
plasma temperature kT$_e~\sim 0.2 $keV, optical depth $\tau$ of cool plasma, photon
index of power-law continuum $\Gamma$, power fraction of power-law continuum f$_{pl}$
and R$_{corona}$. We fixed the M$_{BH}$ at $2\times10^{10}~M_{\odot}$
\citep{2016ApJ...819L..37V,2018MNRAS.473.1145K}, D$_L$ at 1677 Mpc, f$_{pl}=0$ 
as the power--law component accounts for the hot Comptonizing component
and we fixed the normalization to unity to get proper flux and luminosity for the source. We tied the power-law photon index to the
photon index of the {\tt optxagnf} model. Rest of the parameters were allowed to vary.
The fit resulted in $\chi^{2}/\nu= 189.5/156$. Further, since {\tt optxagnf} can
describe hard X-ray power-law continuum, we varied the parameter f$_{pl}$ after
removing analytical {\tt powerlaw} model. This resulted in $\chi^{2}/\nu= 189.2/156$
with best-fit model parameters listed in Table~\ref{table1} and the plot in the top and
bottom panels of Fig.~\ref{fig2}~(e), respectively. Additionally, as per other
the claims of geometrical models, we also tested M$_{BH}$ of $\sim10^{8}~M_{\odot}$.
However, it resulted in a super Eddington accretion rate of about 1.3 in Eddington units, contrary to the 
expectation of BLLs.

\setlength\tabcolsep{1.0pt}
\begin{table}
\fontsize{7.6}{5.5}\selectfont
\centering 
\caption{Best-fitting parameters of models used to fit the 0.3-10~\kev~and the UV/Optical bands jointly for 2015~observations. Model 1 : Nth.+Relxill+diskbb; Model 2 : Optxagnf. `t` stands for tied parameter in the fit.}
\label{table2} 
\begin{tabular}{lcccccc}
\hline 
Model component        &\multicolumn{1}{c}{Model 1} & \multicolumn{1}{c}{~~~~~~~~~~~~~~~~~~~~Model 2}   \\
\hline 
Reddening &$0.13\pm0.03$                             &Reddening & $0.16_{-0.04}^{+0.03}$\\ 
Photon index ($\Gamma$) &$2.22\pm0.08$               &Photon index ($\Gamma$) & $1.87_{-0.03}^{+0.06}$\\ 
$\frac{kT_{bb}(nth)}{eV}$&$2.03_{-0.03}^{+0.09}$ &--&-- \\
Norm.(nth) ($10^{-3}$)&$0.86_{-0.03}^{+0.05}$       &Acc. rate ($\frac{\rm L}{\rm L_{edd}}$)   & $0.040_{-0.004}^{+0.003}$  \\
Index1   ($q1$) &$7.2_{-1.3}^{+1.7}$                &Coronal radius ($r_{g}$)    &$5.7_{-0.9}^{+1.9}$   \\
Index2 ($q2$)    &$4.4_{-1.1}^{+1.7}$               &Plasma temp. ($\frac{\rm kT_e}{\rm keV})$  & $0.47_{-0.12}^{+0.03}$  \\
Photon index ($\Gamma$)&$2.22\pm0.08$ (t)           &Optical depth ($\tau$)     & $9.0_{-0.3}^{+2.0}$  \\
log($\frac{\rm Ionization~par.}{\rm erg~cm~s^{-1}}$) &$2.3_{-0.4}^{+0.1}$  &Frac. power ($f_{pl}$)     & $0.12_{-0.02}^{+0.05}$ \\
Inner radius ($r_{g}$)&$1.6\pm0.3$        &--& --\\
Spin ($a$)  &$0.99_{-0.04}^{+0.01}$ &Spin ($a$)  &$0.996_{-0.006}^{+0.002}$ \\
Break radius  ($r_{g}$) &$4.2_{-1.9}^{+1.7}$    &--&-- \\
Norm.(refl) ($10^{-5}$)&$2.7_{-1.3}^{+1.1}$          &&-- \\
$\frac{kT_{in}(disk)}{eV}$&$2.03_{-0.03}^{+0.09}$ (t) &--&-- \\
Norm.(disk) ($10^{11}$)&$3.20\pm0.01$ &--&-- \\
Stat. ($\chi^{2}/\nu$)&194/156                       &Stat. ($\chi^{2}/\nu$)&200.9/160\\
\hline  
\end{tabular}
\end{table}

Application of these models to the May 2018 0.3-10~\kev ~data did not improve the fit
statistics with respect to a simple {\tt powerlaw}, as can be seen from results
in Table \ref{table1} and, neither the observation show any strong
soft X-ray excess (see Fig.~\ref{fig1}).

\subsubsection{X-ray to UV/Optical Emission}
Statistically, log-parabolic emission ({\tt logpar}), blurred reflection ({\tt relxill}) and cool Comptonization
({\tt optxagnf}) models describe the soft X-ray excess equally well. To distinguish between
these models, we used the UVOT data from \swift{}~UVOT (MJD=57149-57150) snapshots, observed
simultaneously with the 2015 \xmm{}~observation. The extrapolation of best-fitting log-parabolic model of X-rays showed strong residuals in UV/optical band. Having corrected from reddening due to our Galaxy and intrinsic to the source, we added {\tt zbremss} model to describe the thermal emission as claimed in studies. We found that the {\tt zbremss} with {\tt logpar} results in the worst statistic (see Fig. \ref{fig3}(a)).   
We then extrapolated the best-fit 
blurred reflection model to UVOT bands and found positive residuals in the low-energy
bands. Since the reflection model does not include disk component required for
optical-UV spectral break, we added  {\tt diskbb} model with the best-fitting parameters
of blurred reflection model. We applied reddening correction due to our Galaxy and
intrinsic to the source. We fitted the UV/Optical and X-ray bands jointly and the
fit resulted in $\chi^{2}/\nu= 189.4/157$. Similarly, we also fitted the UV/Optical/X-ray
bands using the disk Comptonization model {\tt optxagnf} which includes the intrinsic disk emission. Having applied reddening correction, we modeled the full band and the fit
resulted in $\chi^{2}/\nu= 201.3/160$. The best-fit model, data and residuals are
shown in Fig.~\ref{fig3}~(b) and ~\ref{fig3}~(c) with parameters in Table~\ref{table2}.

\begin{figure}[h]
  \includegraphics[scale=0.38,angle=0.0]{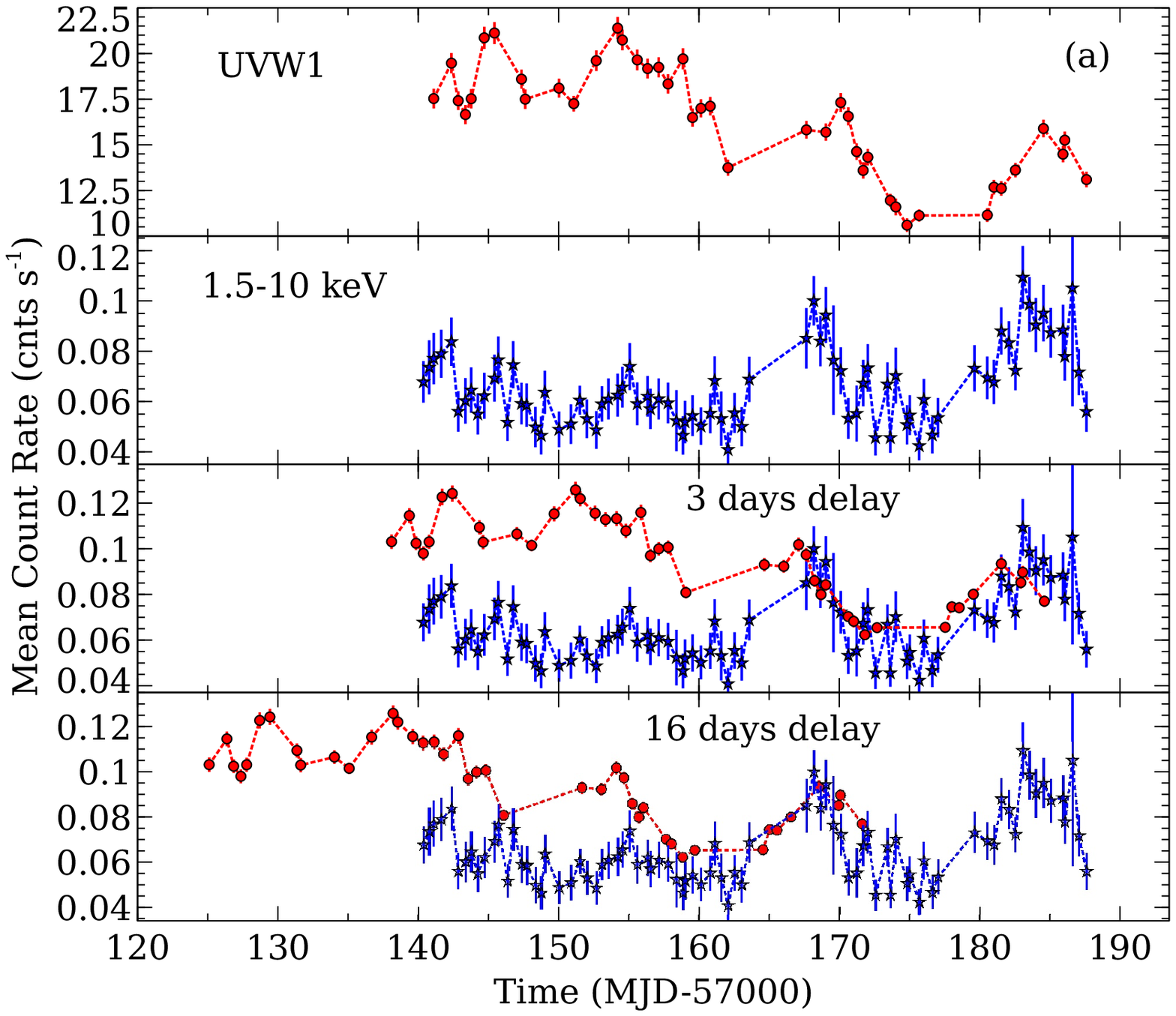}
  \includegraphics[scale=0.43,angle=0.0]{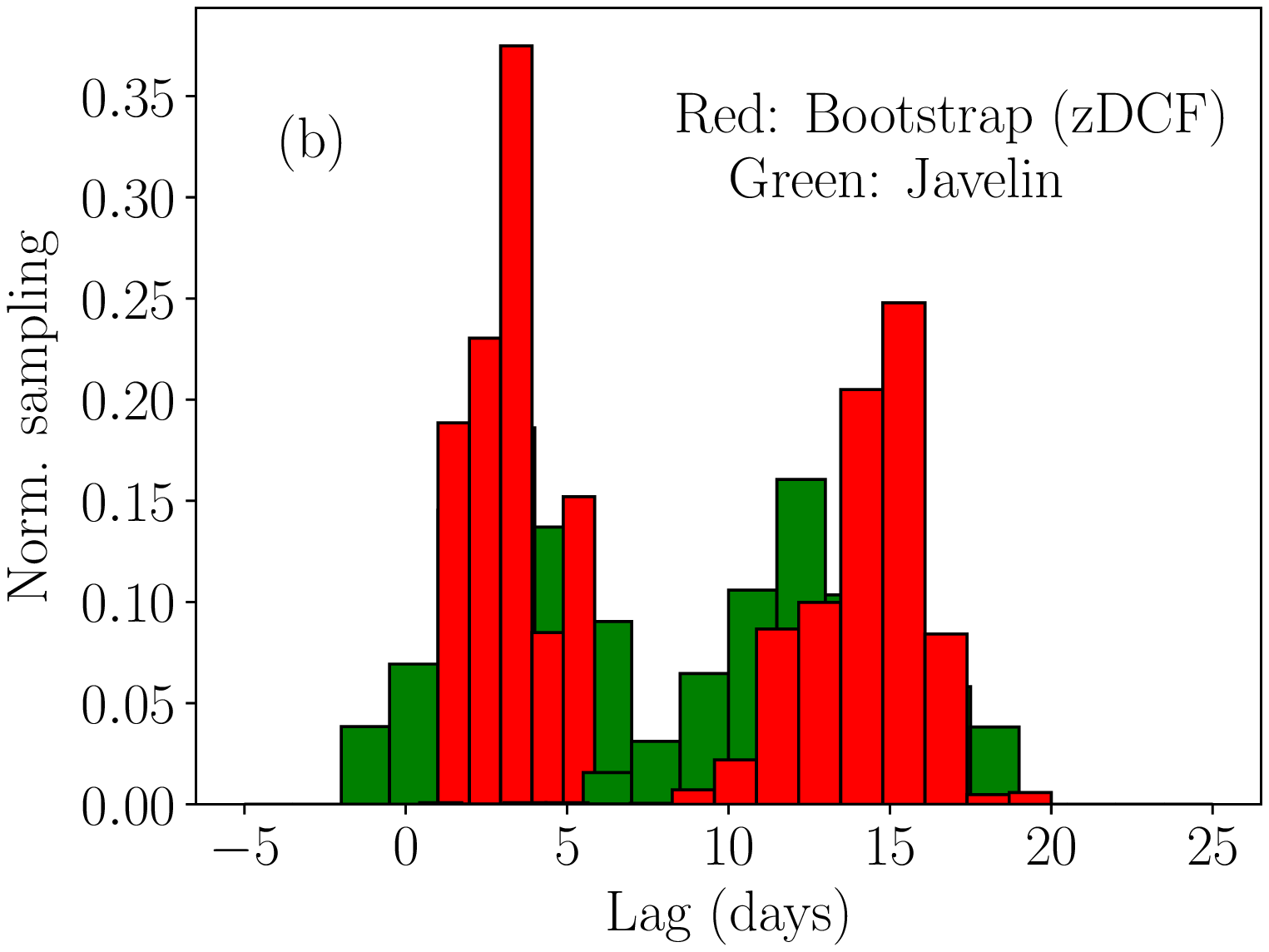}
  \includegraphics[scale=0.37,angle=0.0]{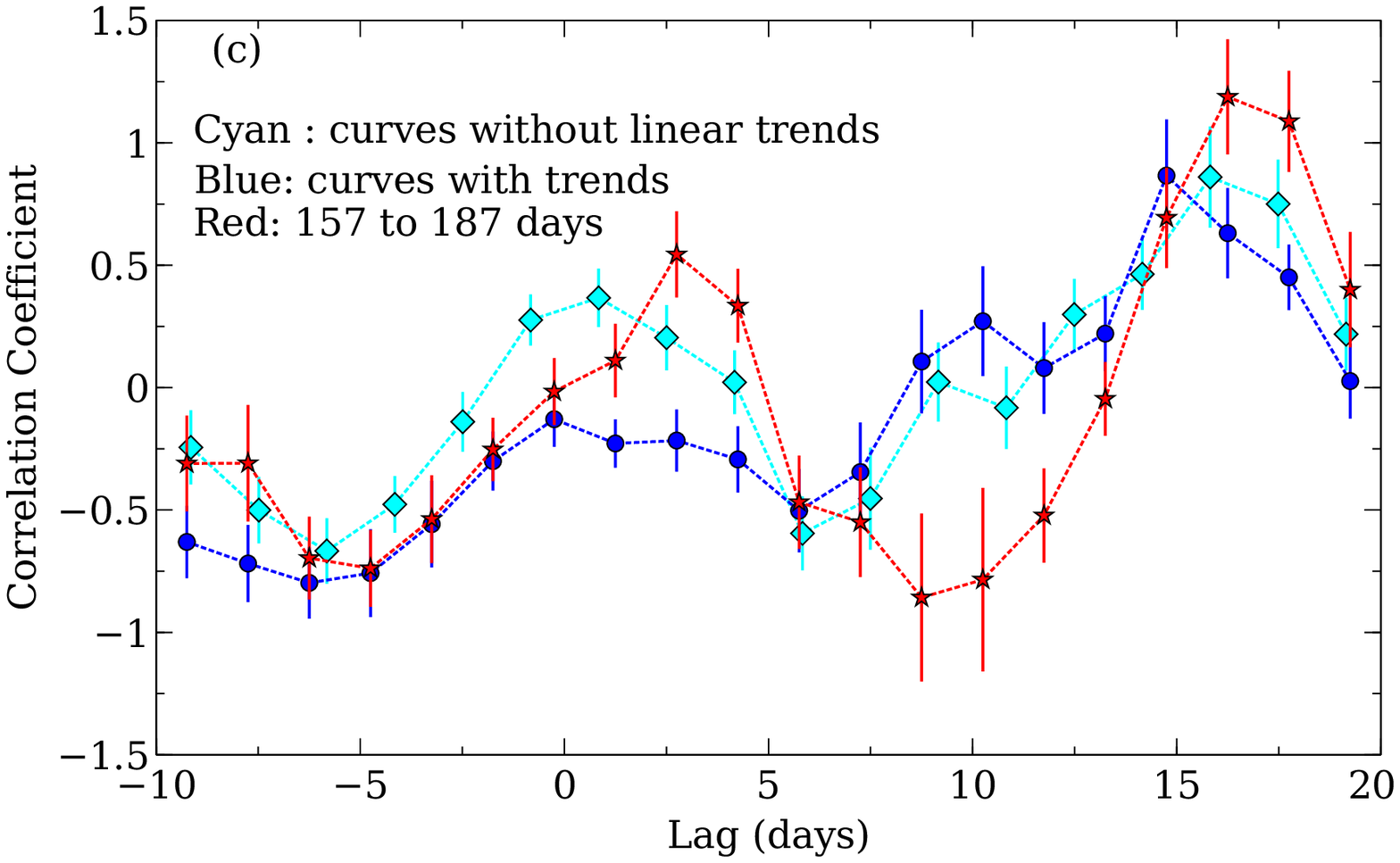}
\caption{{(a)}: Two upper panels show the light curves for UVW1 and
1.5-10 keV bands while the two lower panels show the UVW1 light curve shifted
by $\sim$3-day and 16-day with respect to the X-ray. Here, the UVW1 count rates are 
divided by a factor of 170 to match the level with X-ray count rates. {(b)}:
Probability distribution of time-delay for UVW1 band with respect to the 1.5-10 keV
from {\tt Javelin} (green) and {\tt ZDCF} (red) codes (see \S\ref{sec:time}). {(c)}: 
Lag results from simple {\tt DCF} method considering different time ranges (red,
blue) and by removing the linear trends from the full light curves (top two panels of
4 (a)).}
\label{fig4}	
\end{figure}

\subsection{Timing Analysis}\label{sec:time}
We examined the timing behavior of OJ~287 2015 May observation to look for
lags between X-ray and UV emission. We first checked \xmm{} UVW1 band vis-a-vis
X-rays and did not find any lag between them, suggesting that UV may not be related
to X-ray within $\sim$ day-long observation. In fact, only the optical data
of the 2018 observation show a hint of marginal variability while rest is
statistically consistent with no variability. We then used \swift{}~-XRT and simultaneous
UV observations taken in UVW1 band obtained on a cadence of about half a day during
MJD=57140.4 to MJD=57187.0, as shown in the upper two panels of Fig. \ref{fig4} (a). Observed
light curves are highly variable and the UV band seems to lag behind the 1.5-10 keV X-rays
around MJD=57170 and afterward. This could be due to reprocessing as on short time
scales OJ 287 normally show simultaneous variability \citep{2018MNRAS.473.1145K,
2013MNRAS.433.2380K}, but has shown lag when an additional competing emission component
is present \citep{2018MNRAS.479.1672K}. In such cases, available light curves to date
can not be used for the analysis due to jet dominated emission.

We used {\tt JAVELIN} code \citep{2011ApJ...735...80Z}
to estimate the lag following the procedures described in \citet{2018MNRAS.474.5351P}
and \citet{2017MNRAS.466.1777P}. We found time-lag of $\sim~3$-day and $\sim~16$-
day for UVW1 compared to X-rays (see Fig.~\ref{fig4} (b)). We cross-checked the
lag results with the z-transformed discrete correlation function ({\tt zDCF}:
\citet{2013arXiv1302.1508A}) applying the bootstrap technique \citep{1998PASP..110..660P}.
 In the bootstrap method, we extracted 10000 realizations of the two light curves from 
the observed light curve pair through Monte Carlo approach by randomizing fluxes
and randomly selecting a subset after excluding 20\%~data points. We then performed
cross-correlation on the extracted pairs using the {\tt zDCF} method, as was done
between the originally observed light curves. This approach is a model-independent
way of accounting for the effects of flux uncertainties and irregular
sampling on the cross-correlation result. The lag results from this are shown in
Fig. \ref{fig4} (b) and as can be seen clearly, the time-lag range agrees with the
one found by the {\tt JAVELIN}. 

 Within the limitations of data reported here, both the $\sim3$-day
and $\sim16$-day lag values are supported by the light curves; shown by plotting a
shifted UV light curve with respect to the X-ray in the bottom panels
of Fig. 4~(a). Further cross-checking with simple Discrete Cross-
Correlation ({\tt DCF}; \citet{1988ApJ...333..646E}), we noticed some discrepancies. The 3-day lag feature 
was missing when the full light curves were used ( top two panels of Fig. \ref{fig4}~(a))
but was recovered when the linear trends were removed from the light curves e.g., \citet{2014MNRAS.444.1469M,2018MNRAS.480.2881M}. The 3-day
lag, however, remains if we consider data after MJD 57157 without even removing the linear
trend from the light curves. These outcomes are shown in Fig.~\ref{fig4}~(c).

 In short, while both the lag values are supported by the data, the $\sim 16$-day
lag is consistently present in all the methods while the $\sim$3-day lag is
recovered after eliminating the linear trend while performing {\tt DCF} analysis. 
Unfortunately, gaps before and after the used light curves and also
the sampling of available data do not allow any further analysis (e.g. significance estimate).
Regardless however there is a clear indication of lag.

\section{Discussion} \label{discussion}
We performed a spectral and temporal study of OJ~287 based on the 2015 and 2018 long 
\xmm~{} observations. Except for a marginal hint of
variability in the 2018 \xmm~optical data, rest is statistically consistent with
no variability within each observation ($\rm \sqrt{variance~ of~ rate} \lesssim$ mean error
in the rate). Spectrally, however, the two observations represent a very different
X-ray spectral state of the source. The 2018 X-ray spectrum shows the most generic
spectral state of the source characterized by a powerlaw spectrum \citep{2001PASJ...53...79I,
2009PASJ...61.1011S,2013MNRAS.433.2380K,2017MNRAS.468..426S} while the 2015 X-ray
spectrum shows strong soft X-ray excess with respect to a powerlaw spectrum below 2.0 keV (figure
\ref{fig1}, \S\ref{subsec:spec}). To best of our knowledge, such (soft X-ray) excesses
-- the focus of our study here, has been reported only once in OJ 287 \citep{2001PASJ...53...79I}. We systematically investigated the emission mechanisms behind the origin of
this excess using models motivated from blazar and normal AGN studies.

\subsection{Blazar based models}

Blazars are known for variability in all the domains of observation. Spectral
changes, as reported in this work, at the low-energy end of the X-ray emission
can physically have multiple origins. In addition to the possibility of an altogether
new emission component \citep[e.g.][]{2018MNRAS.473.1145K,2018MNRAS.479.1672K}, in
the general scheme of blazar emission scenario, an appropriate overlap of synchrotron
and SSC component can mimic a variety of phenomenological spectral shapes. 
Literature records on OJ 287 show only one instance of a similar spectral state in the
1994 ASCA observation. A spectral study by \citet{1997PASJ...49..631I} reported a power-law
photon spectral index of $\Gamma \sim 1.67$. However, a careful re-analysis
by \citet{2001PASJ...53...79I} found that a broken-powerlaw spectrum with a break
at 2 keV describe the data statistically better. Further, the
spectral index below 2 keV was consistent with the optical-UV spectrum and hence,
they attributed the soft-excess to the ``synchrotron soft tail''.

\begin{figure}
\includegraphics{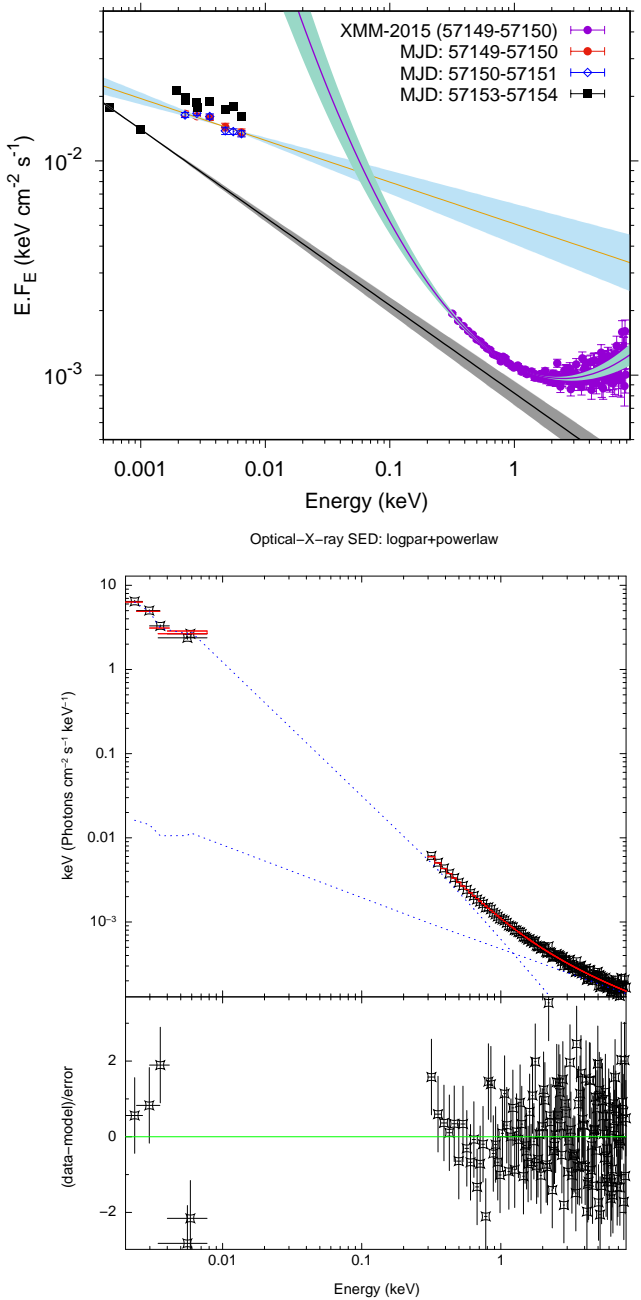}
\includegraphics[scale=0.42]{fig5b.eps}
\caption{Top: NIR to X-ray SEDs of OJ 287 around 2015 \xmm{} observation. The solid
curves within the shaded regions are the best fit logparabola and powerlaw model
to the X-ray and NIR/optical-UV data respectively while shaded area represent their
$1\sigma$ range bounded by error in spectral indices only.  {\it Bottom:} Best-fitted
logparabola+powerlaw model to the optical to X-ray emission. The dotted curves are
the individual model components (optical to soft-X-ray: logparabola) while the red
curve is the sum of the two components.}
\label{fig5}
\end{figure}

In the current case, we followed a
flexible approach and systematically investigated by using with both the possible
phenomenological spectral shapes -- logparabola and broken-powerlaw models. This allows
to capture additional contributions \citep[e.g.][]{2018MNRAS.479.1672K,
2001PASJ...53...79I}. Of the two, we found that a logparabola model provides
statistically acceptable description of the 2015 EPIC-pn data (ref. Table \ref{table1}).
A logparabola spectrum within blazar emission scenarios can simply arise from an
appropriate combination of the high-energy end of a simple powerlaw synchrotron spectrum
or its steeply declining part with the rising part of the SSC emission (ref.
Fig. \ref{fig5}, bottom panel). A look at the NIR and optical SEDs around
the 2015 observation, as shown in the top plot of Fig. \ref{fig5}, clearly show that
the soft X-ray excess lies above the simple powerlaw extrapolation of the NIR data
but below the optical-UV data points. Noting that in the most generic spectral state
of OJ 287 the optical-UV data simply lies on a power-law (log-parabola) extension
of NIR data, the SEDs around 2015 observations suggest two possibilities in the
present context --
{\it CASE-A}:   synchrotron spectrum associated with NIR data points extending
to X-rays with a powerlaw or steeply declining tail (Fig. \ref{fig5}
grey band). {\it CASE-B}: optical-UV being synchrotron with a smoothly declining
tail causing the soft X-ray excess (Fig. \ref{fig5} bottom plot). Below we
systematically look into these two possibilities.

\subparagraph*{ CASE-A:}
In this case, the optical-UV data remains unexplained, suggesting additional
broadband emission component. Attributing NIR-optical break to accretion
disk emission as suggested in \citet{2018MNRAS.473.1145K}, the combined emission still
failed to reproduce the UV emission \citep[see Fig. 6 in][also
\citet{2019BHCB}]{2018MNRAS.473.1145K}. Thus, though this interpretation could
provide a viable explanation for soft X-ray excess, the UV data remain unexplained.

\subparagraph*{ CASE-B}
As shown in Fig. \ref{fig5} bottom plot, this scenario successfully reproduces the
optical/UV to soft X-ray emission by using logparabola model (logpar) when
combined with a powerlaw representing the X-ray emission above the soft X-ray band.
 The combined logparabola plus powerlaw provide an acceptable
fit to the data ($\chi^2/\nu \simeq 1.2$). 
Furthermore, the resulting {\it powerlaw}
index of $\sim$1.6 for X-ray emission is also consistent with the general X-ray
spectra of the source. However, this fails to explain the two NIR data points
unless the synchrotron peak of its broadband SED which normally peaks at NIR
\citep[KJ bands; e.g.][]{2009PASJ...61.1011S,2013MNRAS.433.2380K} has shifted to
optical energies, making
NIR data to be part of spectrum before the peak of synchrotron emission. But NIR-optical
SED comparison with 2009 SED does not support such shift \citep[Fig. 2, ][]
{2019BHCB}. Further, even smoothing of the low-energy
end to match one of the NIR data on the basis that low-energy hump peaks around NIR
bands \citep[e.g.][]{2001PASJ...53...79I,2013MNRAS.433.2380K,2018MNRAS.473.1145K}
leaves the other NIR data points unexplained. It should further be noted that this is
not a one odd observational data as the NIR-optical SED trend has been like this
since May 2013
(MJD 54639) as reported by \citet{2018MNRAS.473.1145K}. Thus, though
phenomenologically {\tt logpar} description is fine for X-rays, it is not consistent
with the broadband emission characteristics of the source during this period,
thereby suggesting some other emission components for the soft X-ray excess.

 Another proposal in the literature is a dominant thermal bremsstrahlung emission
for the $\sim$ 12-yr quasi-periodic optical outbursts from a thermal gas of
temperature $\sim 3\times10^5$ K \citep{2016ApJ...819L..37V, 2012MNRAS.427...77V}.
However, this temperature corresponds to $\sim$ 25 eV, too small to produce the
observed excess in 0.1 -- 2 keV (Fig. \ref{fig3}a). Considering this scenario
and keeping the temperature free during the fit, we found an
statistically acceptable fit with 25 keV plasma. However, this is unphysical 
as it dominates the high-energy end of the X-ray emission and is contradictory
with previous studies and the general X-ray spectral profile of the source being
a powerlaw. Combination of logparabola (for synchrotron and its high energy
tail), thermal bremsstrahlung, and powerlaw (for SSC) to optical to X-ray data
resulted in a very low plasma temperature ($\sim0.1$ eV), making the bremsstrahlung
ineffective with resulting scenario similar to a powerlaw plus logparabola which,
as argued above, are in tension with NIR-optical break.

\subsection{Radio-quiet AGN/disk based models}

The claim of NIR-optical break as accretion disk emission of a $\sim 10^{10}~\rm
M_{\odot}$ SMBH and its presence between May 2013 (MJD 56439) till May 2016 suggests
disk-based soft X-ray excess origin as in radio-quiet AGN as potential
candidates. We, therefore, investigated
this possibility with two of the AGN disk dominant models:  cool Comptonization
({\it optxagn}) and blurred reflection, argued for the soft X-ray excess often
observed in Seyfert galaxies; e.g. Mrk~509 \citep{2011A&A...534A..39M}, 1H~0707--495
\citep{2009Natur.459..540F}, II~Zw~177 \citep{2016MNRAS.457..875P}, ESO~113--G010
\citep{2013ApJ...764L...9C}. We found that both cool Comptonization and blurred
reflection plus disk describes the data well (like the logparabola model) and
are equally acceptable statistically (see Table \ref{table1} and \S\ref{dataAn}).

In the  cool Comptonization scenario, the best-fit suggests the observed
soft excess is due to inverse Compton scattering of seed photons flux from the disk
($f_{PL}\sim 0.1$ , see table \ref{table1}) in the cool Comptonizing plasma
(kT$_{e}\sim$0.4 keV and $\tau\sim$10 in this work). The derived accretion rate for
observed soft X-ray excess was found to be $\sim$10\%~of Eddington unit. Such a high
accretion rate for prominent soft X-ray excess has been seen in a number of radio-loud
narrow line Seyfert type 1 (RLNLS1) AGNs i.e., 1H~0323$+$342 \citep{2018MNRAS.479.2464G}.
The temperature and optical depth of the cool plasma embedded in the inner region of the
accretion disk are inferred to be $kT_{e}\sim0.5$ keV and $\tau\sim 10$, respectively,
for the soft X-ray excess in 2015 May observation. Such type of cool plasma has
been found in RLNLS1 galaxies i.e., PMN~J0948$+$0022: \citep{2014MNRAS.438.3521D}. The
flux observed for soft excess in the 0.3-2 keV band was found to be $\sim 3\times 10^{-12}\rm~erg~s^{-1}~cm^{-2}$
which is comparable to which is claimed in a RLNLS1 PMN~J0948$+$0022. Thus, the BL-Lac object OJ~287 behaves like a radio loud narrow-line Seyfert galaxies in this particular observation. 

Additionally, since SMBH mass is one of the parameter in the cool Comptoniozation
scenario, we also checked it by fitting first a SMBH mass of $\sim 2\times 10^{10}~M_{\odot}$,
as suggested by NIR-optical break and also in the disk-impact binary SMBH model.
This resulted in a accreting rate of $\sim 0.04$ in Eddington units. Fit with an SMBH
mass of $\sim 1\times 10^{8}~M_{\odot}$ as argued by jet precession based models,
on the other hand, resulted in a super Eddington accretion rate $\sim$ 1.3, contrary
to the expectation for BL Lac objects. Thus, the model too supports a very massive SMBH
mass as claimed in the binary SMBH and also from the NIR-optical spectral break.
It should, however, be noted that central engine mass is not a true discriminator
for the two classes of models suggested for $\sim 12$-yr QPO as in the geometrical class
of models the central engine mass is not connected directly with the model parameters
and is inferred based on other observations, unlike the case of disk-impact
binary SMBH model.

In case  of X-ray reflection under the lamppost geometry, the blurred
reflection is very intense and strong (ref. table \ref{table1}) close to the inner
edge of the accretion disk. The emissivity pattern is not uniform and it changes
from inner radius to a break radius $R_{br}\sim 4r_g$ (inner emissivity index$\sim7$
and outer emissivity index $\sim4$). Thus, the strong soft excess is
likely due to the strong light bending in the vicinity of the central SMBH. 
The best known proxy for the blurred reflection is the broad iron-K$\alpha$
emission line near 6 keV \citep{1995Natur.375..659T}. However, the Fe-K$\alpha$ emission
line is absent in the 2015 observation, and in fact never been detected in OJ 287
or any BLL to best of our knowledge. The fit suggests an intense smearing for
blurred reflection, too strong for Fe-K$\alpha$ emission line to be seen in the data
 (see blue dashed line for blurred reflection in Fig. 3(b)). In this scenario,
a likely possibility is that the disk may be illuminated by the base of the jet
\citep{2018MNRAS.473.3584P,2014MNRAS.444.1469M,2018MNRAS.480.2881M}. We found
clear indications of lagging of UVW1 band emission with respect to the hard X-ray emission
($\sim 3$ and $\sim 16$ days, see Fig. \ref{fig4}). Such lags favor the X-ray
reprocessing scenario at accretion disk and have been reported in many AGNs where
UV is found to be lagging behind X-ray emission as expected in the reprocessing scenario
\citep[\eg][]{2017MNRAS.464.3194B, 2018MNRAS.474.5351P, 2014MNRAS.444.1469M,2018MNRAS.480.2881M}. Additional support
for this comes from the general variability trend of OJ 287 where multi-wavelength
variations are normally simultaneous on short timescales \citep{2018MNRAS.473.1145K,
2013MNRAS.433.2380K} with lag reported only when an additional emission component
was competing with its general emission \citep{2018MNRAS.479.1672K}.
 
 The best-fit reflection+disk model in the optical/UV/X-ray bands suggests a inner disk temperature of $\sim$
2 eV (ref. table \ref{table2}). We used theoretical
temperature profile  ($T(r) \sim6.3\times10^5(\frac{\dot M_{E}}{M_{8}})^{0.25}
(\frac{r}{R_s})^{-0.75}$, where $\dot M_{E}$, $M_{8}$, $R_S$ and $r$ are accretion
rate in Eddington units, black
hole mass in $10^8~M_{\odot}$, Schwarzschild radius and disk radius from the centre,
 respectively) with the best-fit parameter to infer the temperature at the inner
edge of the disk. This provided in a temperature of about {$\sim$ 2.8 eV}, similar to
the one inferred from the X-ray/UV/optical modeling, further supporting the
disk-impact binary SMBH scenario. Further, the normalization of 
multi-color blackbody model {i.e. \tt diskbb} is a function of inner
radius of the accretion disk and the luminosity distance along with the
inclination of the source. Using inner radius $R_{in}=1.6~r_g$, mass of
the black hole $M_{BH}=2\times10^{10}~M_{\odot}$, inclination $i=3$
degree and luminosity distance 1652.08 Mpc, we derived the normalization
value to be $8.2\times10^{10}$. This is similar with the 
best-fit value listed in Table~\ref{table2}. Thus, both the observed inner disk temperature and the normalization
are in agreement in support for the binary black hole
system with a super heavy super massive black hole at the centre.

Both the AGN disk-based models
suggest a maximally rotating SMBH, contrary to the tightly constrained spin value
of $\sim 0.30$ claimed by \citet{2016ApJ...819L..37V}. We tested blurred reflection
by fixing the spin parameter at 0.30 and the fit-statistic was marginally
disfavored ($\chi^{2}_\nu\sim$1.5). This marginal change for a large change
in the value of spin suggests that current data are not sufficient to 
constrain the spin and/or a detailed comparative study is required based
on the theoretical premise of the model.

\section{Summary and Conclusion}\label{Conclusion}
 We performed spectral analysis of the two yet unstudied \xmm~{} observations
of OJ 287 performed in 2015 and 2018 respectively. Temporally, both the data are statistically 
consistent with non-variable but are spectrally very different. We found that while
the 2018 data represents the typical (most generic) X-ray spectral state of the source
characterized by a powerlaw spectrum, the 2015 data show very strong soft X-ray excess.
The excess lies above the simple power-law extrapolation of the NIR data points but
below the best-fit power-law extrapolation of the optical-UV data points. We systematically
explored the physical process behind the spectral shape vis-a-vis consistency with
known/established observational properties of OJ 287 as listed below.

\begin{itemize}
 \item For the X-ray spectrum only, a simple log-parabola model describes the 2015 spectral
 state statistically well and can be generated with an appropriate overlap of synchrotron
 tail extended to X-ray energies and the SSC spectrum. However, this interpretation is in conflict with the quasi-simultaneous NIR to optical spectrum of the source. 
 
 \item Additionally adding a thermal bremsstrahlung emission from a plasma of
 temperature 25 keV with logparabola also provides an acceptable statistical fit to
 the X-ray data but is inconsistent with the optical spectrum as well as the general X-ray
 spectral properties of the source.
 
 \item Accretion disk-based models: reflection and cool Comptonization (Table \ref{table1})
 with an intrinsic powerlaw component describes 2015 optical to X-ray spectrum
 statistically well and is consistent with the general spectral characteristics of
 OJ 287. Timing analysis indicates a lag of UV emission with respect
 to X-rays (\S\ref{sec:time}), favoring reflection model. Additionally, these models also favor
 a heavy SMBH of mass $\sim 10^{10}~M_\odot$ for OJ 287, as has been argued by 
 \citet{1988ApJ...325..628S} and \citet{1996ApJ...460..207L} in interpreting the
 $\sim$12-yr optical QPO in a binary SMBH framework.
 
Further, the appearance of the soft excess during 2015 and its absence in 2018 is 
 consistent with the presence of accretion-disk signature (NIR-optical break) between
 May 2013 to November 2016. Based on these considerations, the soft X-ray excess and
 UV emission appear to be primarily a result of \it{reflection phenomena}.
 
\end{itemize}

\section{Acknowledgement}
Authors are grateful to acknowledge the anonymous referee for his/her thoughtful 
suggestions and comments which improved the manuscript. 
MP thanks the financial support of UGC, India program through DSKPDF fellowship
(grant no.~BSR/2017-2018/PH/0111). MP is also grateful for support of Prof. M.
Sami at the Centre for Theoretical Physics, Jamia Millia Islamia, New Delhi. 
PK acknowledge funding from FAPESP (grant no. 2015/13933-0). This research
has made use of archival data of \xmm{}  observatory, an ESA science mission directly
funded by ESA Member States and NASA by the NASA Goddard Space Flight Center (GSFC).
 This research has also made use of the XRT Data Analysis Software (XRTDAS)
developed under the responsibility of the ASI Science Data Center (ASDC), Italy. 

\vspace{5mm}
\facilities{Swift (XRT and UVOT), XMM Newton}

\software{HEASOFT (\url{https://heasarc.gsfc.nasa.gov/docs/software/heasoft/}),
Gnuplot (version: 5.0; \url{http://www.gnuplot.info/})}

\bibliographystyle{aasjournal} 
\bibliography{refs8.bib}

\begin{thebibliography}{}
\expandafter\ifx\csname natexlab\endcsname\relax\def\natexlab#1{#1}\fi
\providecommand{\url}[1]{\href{#1}{#1}}
\providecommand{\dodoi}[1]{doi:~\href{http://doi.org/#1}{\nolinkurl{#1}}}
\providecommand{\doeprint}[1]{\href{http://ascl.net/#1}{\nolinkurl{http://ascl.net/#1}}}
\providecommand{\doarXiv}[1]{\href{https://arxiv.org/abs/#1}{\nolinkurl{https://arxiv.org/abs/#1}}}

\bibitem[{{Abdo} {et~al.}(2010){Abdo}, {Ackermann}, {Agudo}, {Ajello}, {Aller},
  {Aller}, {Angelakis}, {Arkharov}, {Axelsson}, {Bach}, {Baldini}, {Ballet},
  {Barbiellini}, {Bastieri}, {Baughman}, {Bechtol}, {Bellazzini}, {Benitez},
  {Berdyugin}, {Berenji}, {Blandford}, {Bloom}, {Boettcher}, {Bonamente},
  {Borgland}, {Bregeon}, {Brez}, {Brigida}, {Bruel}, {Burnett}, {Burrows},
  {Buson}, {Caliandro}, {Calzoletti}, {Cameron}, {Capalbi}, {Caraveo},
  {Carosati}, {Casandjian}, {Cavazzuti}, {Cecchi}, {{\c{C}}elik}, {Charles},
  {Chaty}, {Chekhtman}, {Chen}, {Chiang}, {Chincarini}, {Ciprini}, {Claus},
  {Cohen- Tanugi}, {Colafrancesco}, {Cominsky}, {Conrad}, {Costamante},
  {Cutini}, {D'ammando}, {Deitrick}, {D'Elia}, {Dermer}, {de Angelis}, {de
  Palma}, {Digel}, {Donnarumma}, {Silva}, {Drell}, {Dubois}, {Dultzin},
  {Dumora}, {Falcone}, {Farnier}, {Favuzzi}, {Fegan}, {Focke}, {Forn{\'e}},
  {Fortin}, {Frailis}, {Fuhrmann}, {Fukazawa}, {Funk}, {Fusco}, {G{\'o}mez},
  {Gargano}, {Gasparrini}, {Gehrels}, {Germani}, {Giebels}, {Giglietto},
  {Giommi}, {Giordano}, {Giuliani}, {Glanzman}, {Godfrey}, {Grenier},
  {Gronwall}, {Grove}, {Guillemot}, {Guiriec}, {Gurwell}, {Hadasch},
  {Hanabata}, {Harding}, {Hayashida}, {Hays}, {Healey}, {Heidt}, {Hiriart},
  {Horan}, {Hoversten}, {Hughes}, {Itoh}, {Jackson}, {J{\'o}hannesson},
  {Johnson}, {Johnson}, {Jorstad}, {Kadler}, {Kamae}, {Katagiri}, {Kataoka},
  {Kawai}, {Kennea}, {Kerr}, {Kimeridze}, {Kn{\"o}dlseder}, {Kocian},
  {Kopatskaya}, {Koptelova}, {Konstantinova}, {Kovalev}, {Kovalev},
  {Kurtanidze}, {Kuss}, {Lande}, {Larionov}, {Latronico}, {Leto}, {Lindfors},
  {Longo}, {Loparco}, {Lott}, {Lovellette}, {Lubrano}, {Madejski}, {Makeev},
  {Marchegiani}, {Marscher}, {Marshall}, {Max-Moerbeck}, {Mazziotta},
  {McConville}, {McEnery}, {Meurer}, {Michelson}, {Mitthumsiri}, {Mizuno},
  {Moiseev}, {Monte}, {Monzani}, {Morselli}, {Moskalenko}, {Murgia},
  {Nestoras}, {Nilsson}, {Nizhelsky}, {Nolan}, {Norris}, {Nuss}, {Ohsugi},
  {Ojha}, {Omodei}, {Orlando}, {Ormes}, {Osborne}, {Ozaki}, {Pacciani},
  {Padovani}, {Pagani}, {Page}, {Paneque}, {Panetta}, {Parent}, {Pasanen},
  {Pavlidou}, {Pelassa}, {Pepe}, {Perri}, {Pesce-Rollins}, {Piranomonte},
  {Piron}, {Pittori}, {Porter}, {Puccetti}, {Rahoui}, {Rain{\`o}}, {Raiteri},
  {Rando}, {Razzano}, {Reimer}, {Reimer}, {Reposeur}, {Richards}, {Ritz},
  {Rochester}, {Rodriguez}, {Romani}, {Ros}, {Roth}, {Roustazadeh}, {Ryde},
  {Sadrozinski}, {Sadun}, {Sanchez}, {Sander}, {Saz Parkinson}, {Scargle},
  {Sellerholm}, {Sgr{\`o}}, {Shaw}, {Sigua}, {Siskind}, {Smith}, {Smith},
  {Spandre}, {Spinelli}, {Starck}, {Stevenson}, {Stratta}, {Strickman},
  {Suson}, {Tajima}, {Takahashi}, {Takahashi}, {Takalo}, {Tanaka}, {Thayer},
  {Thayer}, {Thompson}, {Tibaldo}, {Torres}, {Tosti}, {Tramacere}, {Uchiyama},
  {Usher}, {Vasileiou}, {Verrecchia}, {Vilchez}, {Villata}, {Vitale}, {Waite},
  {Wang}, {Winer}, {Wood}, {Ylinen}, {Zensus}, {Zhekanis}, \&
  {Ziegler}}]{2010ApJ...716...30A}
{Abdo}, A.~A., {Ackermann}, M., {Agudo}, I., {et~al.} 2010, ApJ, 716, 30,
  \dodoi{10.1088/0004-637X/716/1/30}

\bibitem[{{Alexander}(2013)}]{2013arXiv1302.1508A}
{Alexander}, T. 2013, arXiv e-prints, arXiv:1302.1508.
\newblock \doarXiv{1302.1508}

\bibitem[{{Arnaud}(1996)}]{1996ASPC..101...17A}
{Arnaud}, K.~A. 1996, in Astronomical Society of the Pacific Conference Series,
  Vol. 101, Astronomical Data Analysis Software and Systems V, ed. G.~H.
  {Jacoby} \& J.~{Barnes}, 17

\bibitem[{{Britzen} {et~al.}(2018){Britzen}, {Fendt}, {Witzel}, {Qian},
  {Pashchenko}, {Kurtanidze}, {Zajacek}, {Martinez}, {Karas}, {Aller}, {Aller},
  {Eckart}, {Nilsson}, {Ar{\'e}valo}, {Cuadra}, {Subroweit}, \&
  {Witzel}}]{2018MNRAS.478.3199B}
{Britzen}, S., {Fendt}, C., {Witzel}, G., {et~al.} 2018, \mnras, 478, 3199,
  \dodoi{10.1093/mnras/sty1026}

\bibitem[{{Buisson} {et~al.}(2017){Buisson}, {Lohfink}, {Alston}, \&
  {Fabian}}]{2017MNRAS.464.3194B}
{Buisson}, D.~J.~K., {Lohfink}, A.~M., {Alston}, W.~N., \& {Fabian}, A.~C.
  2017, \mnras, 464, 3194, \dodoi{10.1093/mnras/stw2486}

\bibitem[{{Cackett} {et~al.}(2013){Cackett}, {Fabian}, {Zogbhi}, {Kara},
  {Reynolds}, \& {Uttley}}]{2013ApJ...764L...9C}
{Cackett}, E.~M., {Fabian}, A.~C., {Zogbhi}, A., {et~al.} 2013, \apjl, 764, L9,
  \dodoi{10.1088/2041-8205/764/1/L9}

\bibitem[{{Crummy} {et~al.}(2006){Crummy}, {Fabian}, {Gallo}, \&
  {Ross}}]{crum2006}
{Crummy}, J., {Fabian}, A.~C., {Gallo}, L., \& {Ross}, R.~R. 2006, \mnras, 365,
  1067, \dodoi{10.1111/j.1365-2966.2005.09844.x}

\bibitem[{{D'Ammando} {et~al.}(2014){D'Ammando}, {Larsson}, {Orienti},
  {Raiteri}, {Angelakis}, {Carrami{\~n}ana}, {Carrasco}, {Drake}, {Fuhrmann},
  \& {Giroletti}}]{2014MNRAS.438.3521D}
{D'Ammando}, F., {Larsson}, J., {Orienti}, M., {et~al.} 2014, \mnras, 438,
  3521, \dodoi{10.1093/mnras/stt2464}

\bibitem[{{Dauser} {et~al.}(2013){Dauser}, {Garcia}, {Wilms}, {B{\"o}ck},
  {Brenneman}, {Falanga}, {Fukumura}, \& {Reynolds}}]{2013MNRAS.430.1694D}
{Dauser}, T., {Garcia}, J., {Wilms}, J., {et~al.} 2013, \mnras, 430, 1694,
  \dodoi{10.1093/mnras/sts710}

\bibitem[{{Dauser} {et~al.}(2010){Dauser}, {Wilms}, {Reynolds}, \&
  {Brenneman}}]{2010MNRAS.409.1534D}
{Dauser}, T., {Wilms}, J., {Reynolds}, C.~S., \& {Brenneman}, L.~W. 2010,
  \mnras, 409, 1534, \dodoi{10.1111/j.1365-2966.2010.17393.x}

\bibitem[{{Dickey} \& {Lockman}(1990)}]{dickey1990}
{Dickey}, J.~M., \& {Lockman}, F.~J. 1990, \araa, 28, 215,
  \dodoi{10.1146/annurev.aa.28.090190.001243}

\bibitem[{{Done} {et~al.}(2012){Done}, {Davis}, {Jin}, {Blaes}, \&
  {Ward}}]{2012MNRAS.420.1848D}
{Done}, C., {Davis}, S.~W., {Jin}, C., {Blaes}, O., \& {Ward}, M. 2012, \mnras,
  420, 1848, \dodoi{10.1111/j.1365-2966.2011.19779.x}

\bibitem[{{Edelson} \& {Krolik}(1988)}]{1988ApJ...333..646E}
{Edelson}, R.~A., \& {Krolik}, J.~H. 1988, \apj, 333, 646,
  \dodoi{10.1086/166773}

\bibitem[{{Fabian} {et~al.}(2009){Fabian}, {Zoghbi}, {Ross}, {Uttley}, {Gallo},
  {Brandt}, {Blustin}, {Boller}, {Caballero-Garcia}, {Larsson}, {Miller},
  {Miniutti}, {Ponti}, {Reis}, {Reynolds}, {Tanaka}, \&
  {Young}}]{2009Natur.459..540F}
{Fabian}, A.~C., {Zoghbi}, A., {Ross}, R.~R., {et~al.} 2009, \nat, 459, 540,
  \dodoi{10.1038/nature08007}

\bibitem[{{Gabriel} {et~al.}(2004){Gabriel}, {Denby}, {Fyfe}, {Hoar}, {Ibarra},
  {Ojero}, {Osborne}, {Saxton}, {Lammers}, \& {Vacanti}}]{2004ASPC..314..759G}
{Gabriel}, C., {Denby}, M., {Fyfe}, D.~J., {et~al.} 2004, in Astronomical
  Society of the Pacific Conference Series, Vol. 314, Astronomical Data
  Analysis Software and Systems (ADASS) XIII, ed. F.~{Ochsenbein}, M.~G.
  {Allen}, \& D.~{Egret}, 759

\bibitem[{{Garc{\'{\i}}a} {et~al.}(2013){Garc{\'{\i}}a}, {Dauser}, {Reynolds},
  {Kallman}, {McClintock}, {Wilms}, \& {Eikmann}}]{2013ApJ...768..146G}
{Garc{\'{\i}}a}, J., {Dauser}, T., {Reynolds}, C.~S., {et~al.} 2013, \apj, 768,
  146, \dodoi{10.1088/0004-637X/768/2/146}

\bibitem[{{Garc{\'{\i}}a} {et~al.}(2011){Garc{\'{\i}}a}, {Kallman}, \&
  {Mushotzky}}]{2011ApJ...731..131G}
{Garc{\'{\i}}a}, J., {Kallman}, T.~R., \& {Mushotzky}, R.~F. 2011, \apj, 731,
  131, \dodoi{10.1088/0004-637X/731/2/131}

\bibitem[{{Garc{\'{\i}}a} {et~al.}(2014){Garc{\'{\i}}a}, {Dauser}, {Lohfink},
  {Kallman}, {Steiner}, {McClintock}, {Brenneman}, {Wilms}, {Eikmann},
  {Reynolds}, \& {Tombesi}}]{2014ApJ...782...76G}
{Garc{\'{\i}}a}, J., {Dauser}, T., {Lohfink}, A., {et~al.} 2014, \apj, 782, 76,
  \dodoi{10.1088/0004-637X/782/2/76}

\bibitem[{{Gaur} {et~al.}(2018){Gaur}, {Mohan}, {Wierzcholska}, \&
  {Gu}}]{2018MNRAS.473.3638G}
{Gaur}, H., {Mohan}, P., {Wierzcholska}, A., \& {Gu}, M. 2018, \mnras, 473,
  3638, \dodoi{10.1093/mnras/stx2553}

\bibitem[{{Ghosh} {et~al.}(2018){Ghosh}, {Dewangan}, {Mallick}, \&
  {Raychaudhuri}}]{2018MNRAS.479.2464G}
{Ghosh}, R., {Dewangan}, G.~C., {Mallick}, L., \& {Raychaudhuri}, B. 2018,
  \mnras, 479, 2464, \dodoi{10.1093/mnras/sty1571}

\bibitem[{{Gierli{\'n}ski} \& {Done}(2004)}]{2004MNRAS.349L...7G}
{Gierli{\'n}ski}, M., \& {Done}, C. 2004, \mnras, 349, L7,
  \dodoi{10.1111/j.1365-2966.2004.07687.x}

\bibitem[{{Goyal} {et~al.}(2018){Goyal}, {Stawarz}, {Zola}, {Marchenko},
  {Soida}, {Nilsson}, {Ciprini}, {Baran}, {Ostrowski}, {Wiita},
  {Gopal-Krishna}, {Siemiginowska}, {Sobolewska}, {Jorstad}, {Marscher},
  {Aller}, {Aller}, {Hovatta}, {Caton}, {Reichart}, {Matsumoto}, {Sadakane},
  {Gazeas}, {Kidger}, {Piirola}, {Jermak}, {Alicavus}, {Baliyan}, {Baransky},
  {Berdyugin}, {Blay}, {Boumis}, {Boyd}, {Bufan}, {Campas Torrent}, {Campos},
  {Carrillo G{\'o}mez}, {Dalessio}, {Debski}, {Dimitrov}, {Drozdz}, {Er},
  {Erdem}, {Escartin P{\'e}rez}, {Fallah Ramazani}, {Filippenko}, {Gafton},
  {Garcia}, {Godunova}, {G{\'o}mez Pinilla}, {Gopinathan}, {Haislip}, {Haque},
  {Harmanen}, {Hudec}, {Hurst}, {Ivarsen}, {Joshi}, {Kagitani}, {Karaman},
  {Karjalainen}, {Kaur}, {Kozie{\l}-Wierzbowska}, {Kuligowska}, {Kundera},
  {Kurowski}, {Kvammen}, {LaCluyze}, {Lee}, {Liakos}, {Lozano de Haro},
  {Moore}, {Mugrauer}, {Naves Nogues}, {Neely}, {Ogloza}, {Okano}, {Pajdosz},
  {Pandey}, {Perri}, {Poyner}, {Provencal}, {Pursimo}, {Raj}, {Rajkumar},
  {Reinthal}, {Reynolds}, {Saario}, {Sadegi}, {Sakanoi}, {Salto Gonz{\'a}lez},
  {Sameer}, {Simon}, {Siwak}, {Schweyer}, {Sold{\'a}n Alfaro}, {Sonbas},
  {Strobl}, {Takalo}, {Tremosa Espasa}, {Valdes}, {Vasylenko}, {Verrecchia},
  {Webb}, {Yoneda}, {Zejmo}, {Zheng}, {Zielinski}, {Janik}, {Chavushyan},
  {Mohammed}, {Cheung}, \& {Giroletti}}]{2018ApJ...863..175G}
{Goyal}, A., {Stawarz}, {\L}., {Zola}, S., {et~al.} 2018, \apj, 863, 175,
  \dodoi{10.3847/1538-4357/aad2de}

\bibitem[{{Gupta} {et~al.}(2016){Gupta}, {Kalita}, {Gaur}, \&
  {Duorah}}]{2016MNRAS.462.1508G}
{Gupta}, A.~C., {Kalita}, N., {Gaur}, H., \& {Duorah}, K. 2016, \mnras, 462,
  1508, \dodoi{10.1093/mnras/stw1667}

\bibitem[{{Gupta} {et~al.}(2017){Gupta}, {Agarwal}, {Mishra}, {Gaur}, {Wiita},
  {Gu}, {Kurtanidze}, {Damljanovic}, {Uemura}, {Semkov}, {Strigachev},
  {Bachev}, {Vince}, {Zhang}, {Villarroel}, {Kushwaha}, {Pandey}, {Abe},
  {Chanishvili}, {Chigladze}, {Fan}, {Hirochi}, {Itoh}, {Kanda}, {Kawabata},
  {Kimeridze}, {Kurtanidze}, {Latev}, {Dimitrova}, {Nakaoka}, {Nikolashvili},
  {Shiki}, {Sigua}, \& {Spassov}}]{2017MNRAS.465.4423G}
{Gupta}, A.~C., {Agarwal}, A., {Mishra}, A., {et~al.} 2017, \mnras, 465, 4423,
  \dodoi{10.1093/mnras/stw3045}

\bibitem[{{Hudec} {et~al.}(2013){Hudec}, {Ba{\v s}ta}, {Pihajoki}, \&
  {Valtonen}}]{2013A&A...559A..20H}
{Hudec}, R., {Ba{\v s}ta}, M., {Pihajoki}, P., \& {Valtonen}, M. 2013, \aap,
  559, A20, \dodoi{10.1051/0004-6361/201219323}

\bibitem[{{Idesawa} {et~al.}(1997){Idesawa}, {Tashiro}, {Makishima}, {Kubo},
  {Otani}, {Fujimoto}, {Kii}, {Makino}, {Takahashi}, {Ueda}, \&
  {Ohashi}}]{1997PASJ...49..631I}
{Idesawa}, E., {Tashiro}, M., {Makishima}, K., {et~al.} 1997, \pasj, 49, 631,
  \dodoi{10.1093/pasj/49.6.631}

\bibitem[{{Isobe} {et~al.}(2001){Isobe}, {Tashiro}, {Sugiho}, \&
  {Makishima}}]{2001PASJ...53...79I}
{Isobe}, N., {Tashiro}, M., {Sugiho}, M., \& {Makishima}, K. 2001, PASJ, 53,
  79, \dodoi{10.1093/pasj/53.1.79}

\bibitem[{{Kapanadze} {et~al.}(2018){Kapanadze}, {Vercellone}, {Romano},
  {Hughes}, {Aller}, {Aller}, {Kapanadze}, \& {Tabagari}}]{2018MNRAS.480..407K}
{Kapanadze}, B., {Vercellone}, S., {Romano}, P., {et~al.} 2018, MNRAS, 480,
  407, \dodoi{10.1093/mnras/sty1803}

\bibitem[{{Kushwaha} {et~al.}(2019){Kushwaha}, {de Gouveia Dal Pino}, {Gupta},
  \& {Wiita}}]{2019BHCB}
{Kushwaha}, P., {de Gouveia Dal Pino}, E.~M., {Gupta}, A.~C., \& {Wiita}, P.~J.
  2019, in Proceeding of Science, Vol. 329, PoS(BHCB2018), 22,
  \dodoi{10.22323/1.329.0022}

\bibitem[{{Kushwaha} {et~al.}(2013){Kushwaha}, {Sahayanathan}, \&
  {Singh}}]{2013MNRAS.433.2380K}
{Kushwaha}, P., {Sahayanathan}, S., \& {Singh}, K.~P. 2013, \mnras, 433, 2380,
  \dodoi{10.1093/mnras/stt904}

\bibitem[{{Kushwaha} {et~al.}(2018{\natexlab{a}}){Kushwaha}, {Gupta}, {Wiita},
  {Gaur}, {de Gouveia Dal Pino}, {Bhagwan}, {Kurtanidze}, {Larionov},
  {Damljanovic}, {Uemura}, {Semkov}, {Strigachev}, {Bachev}, {Vince}, {Gu},
  {Zhang}, {Abe}, {Agarwal}, {Borman}, {Fan}, {Grishina}, {Hirochi}, {Itoh},
  {Kawabata}, {Kopatskaya}, {Kurtanidze}, {Larionova}, {Larionova}, {Mishra},
  {Morozova}, {Nakaoka}, {Nikolashvili}, {Savchenko}, {Troitskaya}, {Troitsky},
  \& {Vasilyev}}]{2018MNRAS.473.1145K}
{Kushwaha}, P., {Gupta}, A.~C., {Wiita}, P.~J., {et~al.} 2018{\natexlab{a}},
  \mnras, 473, 1145, \dodoi{10.1093/mnras/stx2394}

\bibitem[{{Kushwaha} {et~al.}(2018{\natexlab{b}}){Kushwaha}, {Gupta}, {Wiita},
  {Pal}, {Gaur}, {de Gouveia Dal Pino}, {Kurtanidze}, {Semkov}, {Damljanovic},
  {Hu}, {Uemura}, {Vince}, {Darriba}, {Gu}, {Bachev}, {Chen}, {Itoh},
  {Kawabata}, {Kurtanidze}, {Nakaoka}, {Nikolashvili}, {Sigua}, {Strigachev},
  \& {Zhang}}]{2018MNRAS.479.1672K}
---. 2018{\natexlab{b}}, \mnras, 479, 1672, \dodoi{10.1093/mnras/sty1499}

\bibitem[{{Lehto} \& {Valtonen}(1996)}]{1996ApJ...460..207L}
{Lehto}, H.~J., \& {Valtonen}, M.~J. 1996, \apj, 460, 207,
  \dodoi{10.1086/176962}

\bibitem[{{Leighly}(1999)}]{1999ApJS..125..317L}
{Leighly}, K.~M. 1999, \apjs, 125, 317, \dodoi{10.1086/313287}

\bibitem[{{McHardy} {et~al.}(2014){McHardy}, {Cameron}, {Dwelly}, {Connolly},
  {Lira}, {Emmanoulopoulos}, {Gelbord}, {Breedt}, {Arevalo}, \&
  {Uttley}}]{2014MNRAS.444.1469M}
{McHardy}, I.~M., {Cameron}, D.~T., {Dwelly}, T., {et~al.} 2014, \mnras, 444,
  1469, \dodoi{10.1093/mnras/stu1636}

\bibitem[{{McHardy} {et~al.}(2018){McHardy}, {Connolly}, {Horne}, {Cackett},
  {Gelbord}, {Peterson}, {Pahari}, {Gehrels}, {Goad}, {Lira}, {Arevalo},
  {Baldi}, {Brandt}, {Breedt}, {Chand}, {Dewangan}, {Done}, {Elvis},
  {Emmanoulopoulos}, {Fausnaugh}, {Kaspi}, {Kochanek}, {Korista}, {Papadakis},
  {Rao}, {Uttley}, {Vestergaard}, \& {Ward}}]{2018MNRAS.480.2881M}
{McHardy}, I.~M., {Connolly}, S.~D., {Horne}, K., {et~al.} 2018, \mnras, 480,
  2881, \dodoi{10.1093/mnras/sty1983}

\bibitem[{{Mehdipour} {et~al.}(2011){Mehdipour}, {Branduardi-Raymont},
  {Kaastra}, {Petrucci}, {Kriss}, {Ponti}, {Blustin}, {Paltani}, {Cappi},
  {Detmers}, \& {Steenbrugge}}]{2011A&A...534A..39M}
{Mehdipour}, M., {Branduardi-Raymont}, G., {Kaastra}, J.~S., {et~al.} 2011,
  \aap, 534, A39, \dodoi{10.1051/0004-6361/201116875}

\bibitem[{{Miniutti} \& {Fabian}(2004)}]{2004MNRAS.349.1435M}
{Miniutti}, G., \& {Fabian}, A.~C. 2004, \mnras, 349, 1435,
  \dodoi{10.1111/j.1365-2966.2004.07611.x}

\bibitem[{{Nilsson} {et~al.}(2010){Nilsson}, {Takalo}, {Lehto}, \&
  {Sillanp{\"a}{\"a}}}]{2010A&A...516A..60N}
{Nilsson}, K., {Takalo}, L.~O., {Lehto}, H.~J., \& {Sillanp{\"a}{\"a}}, A.
  2010, \aap, 516, A60, \dodoi{10.1051/0004-6361/201014198}

\bibitem[{{Pal} {et~al.}(2017){Pal}, {Dewangan}, {Connolly}, \&
  {Misra}}]{2017MNRAS.466.1777P}
{Pal}, M., {Dewangan}, G.~C., {Connolly}, S.~D., \& {Misra}, R. 2017, \mnras,
  466, 1777, \dodoi{10.1093/mnras/stw3173}

\bibitem[{{Pal} {et~al.}(2018){Pal}, {Dewangan}, {Kembhavi}, {Misra}, \&
  {Naik}}]{2018MNRAS.473.3584P}
{Pal}, M., {Dewangan}, G.~C., {Kembhavi}, A.~K., {Misra}, R., \& {Naik}, S.
  2018, \mnras, 473, 3584, \dodoi{10.1093/mnras/stx2608}

\bibitem[{{Pal} {et~al.}(2016){Pal}, {Dewangan}, {Misra}, \&
  {Pawar}}]{2016MNRAS.457..875P}
{Pal}, M., {Dewangan}, G.~C., {Misra}, R., \& {Pawar}, P.~K. 2016, \mnras, 457,
  875, \dodoi{10.1093/mnras/stw009}

\bibitem[{{Pal} \& {Naik}(2018)}]{2018MNRAS.474.5351P}
{Pal}, M., \& {Naik}, S. 2018, \mnras, 474, 5351, \dodoi{10.1093/mnras/stx3103}

\bibitem[{{Peterson} {et~al.}(1998){Peterson}, {Wanders}, {Horne}, {Collier},
  {Alexander}, {Kaspi}, \& {Maoz}}]{1998PASP..110..660P}
{Peterson}, B.~M., {Wanders}, I., {Horne}, K., {et~al.} 1998, \pasp, 110, 660,
  \dodoi{10.1086/316177}

\bibitem[{{Qian}(2018)}]{2018arXiv181111514Q}
{Qian}, S. 2018, arXiv e-prints, arXiv:1811.11514.
\newblock \doarXiv{1811.11514}

\bibitem[{{Seta} {et~al.}(2009){Seta}, {Isobe}, {Tashiro}, {Yaji}, {Arai},
  {Fukuhara}, {Kohno}, {Nakanishi}, {Sasada}, {Shimajiri}, {Tosaki}, {Uemura},
  {Anderhub}, {Antonelli}, {Antoranz}, {Backes}, {Baixeras}, {Balestra},
  {Barrio}, {Bastieri}, {Becerra Gonz{\'a}lez}, {Becker}, {Bednarek}, {Berger},
  {Bernardini}, {Biland}, {Bock}, {Bonnoli}, {Bordas}, {Borla Tridon},
  {Bosch-Ramon}, {Bose}, {Braun}, {Bretz}, {Britvitch}, {Camara}, {Carmona},
  {Commichau}, {Contreras}, {Cortina}, {Costado Dios}, {Covino}, {Curtef},
  {Dazzi}, {de Angelis}, {de Cea Del Pozo}, {de Los Reyes}, {de Lotto}, {de
  Maria}, {de Sabata}, {Delgado M{\'e}ndez}, {Dom{\'\i}nguez}, {Dorner},
  {Doro}, {Elsaesser}, {Errando}, {Ferenc}, {Fern{\'a}ndez}, {Firpo},
  {Fonseca}, {Font}, {Galante}, {Garc{\'\i}a L{\'o}pez}, {Garczarczyk}, {Gaug},
  {Goebel}, {Hadasch}, {Hayashida}, {Herrero}, {Hildebrand},
  {H{\"o}hne-M{\"o}nch}, {Hose}, {Hsu}, {Jogler}, {Kranich}, {La Barbera},
  {Laille}, {Leonardo}, {Lindfors}, {Lombardi}, {Longo}, {L{\'o}pez}, {Lorenz},
  {Majumdar}, {Maneva}, {Mankuzhiyil}, {Mannheim}, {Maraschi}, {Mariotti},
  {Mart{\'\i}nez}, {Mazin}, {Meucci}, {Meyer}, {Miguel Miranda}, {Mirzoyan},
  {Miyamoto}, {Mold{\'o}n}, {Moles}, {Moralejo}, {Nieto}, {Nilsson},
  {Ninkovic}, {Otte}, {Oya}, {Paoletti}, {Paredes}, {Pasanen}, {Pascoli},
  {Pauss}, {Pegna}, {Perez-Torres}, {Persic}, {Peruzzo}, {Prada}, {Prandini},
  {Puchades}, {Reichardt}, {Rhode}, {Rib{\'o}}, {Rico}, {Rissi}, {Robert},
  {R{\"u}gamer}, {Saggion}, {Saito}, {Salvati}, {S{\'a}nchez-Conde},
  {Satalecka}, {Scalzotto}, {Scapin}, {Schweizer}, {Shayduk}, {Shore}, {Sidro},
  {Sierpowska-Bartosik}, {Sillanp{\"a}{\"a}}, {Sitarek}, {Sobczynska},
  {Spanier}, {Stamerra}, {Stark Schneebeli}, {Takalo}, {Tavecchio}, {Temnikov},
  {Tescaro}, {Teshima}, {Tluczykont}, {Torres}, {Turini}, {Vankov}, {Wagner},
  {Wittek}, {Zabalza}, {Zandanel}, {Zanin}, \&
  {Zapatero}}]{2009PASJ...61.1011S}
{Seta}, H., {Isobe}, N., {Tashiro}, M.~S., {et~al.} 2009, Publications of the
  Astronomical Society of Japan, 61, 1011, \dodoi{10.1093/pasj/61.5.1011}

\bibitem[{{Shi} {et~al.}(2007){Shi}, {Liu}, \& {Song}}]{2007Ap&SS.310...59S}
{Shi}, W., {Liu}, X., \& {Song}, H. 2007, Ap\&SS, 310, 59,
  \dodoi{10.1007/s10509-007-9413-z}

\bibitem[{{Siejkowski} \& {Wierzcholska}(2017)}]{2017MNRAS.468..426S}
{Siejkowski}, H., \& {Wierzcholska}, A. 2017, \mnras, 468, 426,
  \dodoi{10.1093/mnras/stx495}

\bibitem[{{Sillanpaa} {et~al.}(1988){Sillanpaa}, {Haarala}, {Valtonen},
  {Sundelius}, \& {Byrd}}]{1988ApJ...325..628S}
{Sillanpaa}, A., {Haarala}, S., {Valtonen}, M.~J., {Sundelius}, B., \& {Byrd},
  G.~G. 1988, \apj, 325, 628, \dodoi{10.1086/166033}

\bibitem[{{Sillanpaa} {et~al.}(1996{\natexlab{a}}){Sillanpaa}, {Takalo},
  {Pursimo}, {Lehto}, {Nilsson}, {Teerikorpi}, {Heinaemaeki}, {Kidger}, {de
  Diego}, {Gonzalez-Perez}, {Rodriguez-Espinosa}, {Mahoney}, {Boltwood},
  {Dultzin-Hacyan}, {Benitez}, {Turner}, {Robertson}, {Honeycut}, {Efimov},
  {Shakhovskoy}, {Charles}, {Schramm}, {Borgeest}, {Linde}, {Weneit}, {Kuehl},
  {Schramm}, {Sadun}, {Grashuis}, {Heidt}, {Wagner}, {Bock}, {Kuemmel},
  {Heines}, {Fiorucci}, {Tosti}, {Ghisellini}, {Raiteri}, {Villata}, {de
  Francesco}, {Bosio}, \& {Latini}}]{1996A&A...305L..17S}
{Sillanpaa}, A., {Takalo}, L.~O., {Pursimo}, T., {et~al.} 1996{\natexlab{a}},
  \aap, 305, L17

\bibitem[{{Sillanpaa} {et~al.}(1996{\natexlab{b}}){Sillanpaa}, {Takalo},
  {Pursimo}, {Nilsson}, {Heinamaki}, {Katajainen}, {Pietila}, {Hanski},
  {Rekola}, {Kidger}, {Boltwood}, {Turner}, {Robertson}, {Honeycut}, {Efimov},
  {Shakhovskoy}, {Fiorucci}, {Tosti}, {Ghisellini}, {Raiteri}, {Villata}, {de
  Francesco}, {Lanteri}, {Chiaberge}, {Peila}, \&
  {Heidt}}]{1996A&A...315L..13S}
---. 1996{\natexlab{b}}, \aap, 315, L13

\bibitem[{{Sitko} \& {Junkkarinen}(1985)}]{1985PASP...97.1158S}
{Sitko}, M.~L., \& {Junkkarinen}, V.~T. 1985, \pasp, 97, 1158,
  \dodoi{10.1086/131679}

\bibitem[{{Stickel} {et~al.}(1989){Stickel}, {Fried}, \&
  {Kuehr}}]{1989A&AS...80..103S}
{Stickel}, M., {Fried}, J.~W., \& {Kuehr}, H. 1989, \aaps, 80, 103

\bibitem[{{Tanaka} {et~al.}(1995){Tanaka}, {Nandra}, {Fabian}, {Inoue},
  {Otani}, {Dotani}, {Hayashida}, {Iwasawa}, {Kii}, {Kunieda}, {Makino}, \&
  {Matsuoka}}]{1995Natur.375..659T}
{Tanaka}, Y., {Nandra}, K., {Fabian}, A.~C., {et~al.} 1995, \nat, 375, 659,
  \dodoi{10.1038/375659a0}

\bibitem[{{Turner} {et~al.}(2001){Turner}, {Abbey}, {Arnaud}, {Balasini},
  {Barbera}, {Belsole}, {Bennie}, {Bernard}, {Bignami}, {Boer}, {Briel},
  {Butler}, {Cara}, {Chabaud}, {Cole}, {Collura}, {Conte}, {Cros}, {Denby},
  {Dhez}, {Di Coco}, {Dowson}, {Ferrando}, {Ghizzardi}, {Gianotti}, {Goodall},
  {Gretton}, {Griffiths}, {Hainaut}, {Hochedez}, {Holland}, {Jourdain},
  {Kendziorra}, {Lagostina}, {Laine}, {La Palombara}, {Lortholary}, {Lumb},
  {Marty}, {Molendi}, {Pigot}, {Poindron}, {Pounds}, {Reeves}, {Reppin},
  {Rothenflug}, {Salvetat}, {Sauvageot}, {Schmitt}, {Sembay}, {Short},
  {Spragg}, {Stephen}, {Str{\"u}der}, {Tiengo}, {Trifoglio}, {Tr{\"u}mper},
  {Vercellone}, {Vigroux}, {Villa}, {Ward}, {Whitehead}, \&
  {Zonca}}]{turner2001}
{Turner}, M.~J.~L., {Abbey}, A., {Arnaud}, M., {et~al.} 2001, \aap, 365, L27,
  \dodoi{10.1051/0004-6361:20000087}

\bibitem[{{Valtonen} {et~al.}(2012){Valtonen}, {Ciprini}, \&
  {Lehto}}]{2012MNRAS.427...77V}
{Valtonen}, M.~J., {Ciprini}, S., \& {Lehto}, H.~J. 2012, \mnras, 427, 77,
  \dodoi{10.1111/j.1365-2966.2012.21861.x}

\bibitem[{{Valtonen} {et~al.}(2016){Valtonen}, {Zola}, {Ciprini}, {Gopakumar},
  {Matsumoto}, {Sadakane}, {Kidger}, {Gazeas}, {Nilsson}, {Berdyugin},
  {Piirola}, {Jermak}, {Baliyan}, {Alicavus}, {Boyd}, {Campas Torrent},
  {Campos}, {Carrillo G{\'o}mez}, {Caton}, {Chavushyan}, {Dalessio}, {Debski},
  {Dimitrov}, {Drozdz}, {Er}, {Erdem}, {Escartin P{\'e}rez}, {Fallah Ramazani},
  {Filippenko}, {Ganesh}, {Garcia}, {G{\'o}mez Pinilla}, {Gopinathan},
  {Haislip}, {Hudec}, {Hurst}, {Ivarsen}, {Jelinek}, {Joshi}, {Kagitani},
  {Kaur}, {Keel}, {LaCluyze}, {Lee}, {Lindfors}, {Lozano de Haro}, {Moore},
  {Mugrauer}, {Naves Nogues}, {Neely}, {Nelson}, {Ogloza}, {Okano}, {Pandey},
  {Perri}, {Pihajoki}, {Poyner}, {Provencal}, {Pursimo}, {Raj}, {Reichart},
  {Reinthal}, {Sadegi}, {Sakanoi}, {Salto Gonz{\'a}lez}, {Sameer}, {Schweyer},
  {Siwak}, {Sold{\'a}n Alfaro}, {Sonbas}, {Steele}, {Stocke}, {Strobl},
  {Takalo}, {Tomov}, {Tremosa Espasa}, {Valdes}, {Valero P{\'e}rez},
  {Verrecchia}, {Webb}, {Yoneda}, {Zejmo}, {Zheng}, {Telting}, {Saario},
  {Reynolds}, {Kvammen}, {Gafton}, {Karjalainen}, {Harmanen}, \&
  {Blay}}]{2016ApJ...819L..37V}
{Valtonen}, M.~J., {Zola}, S., {Ciprini}, S., {et~al.} 2016, \apjl, 819, L37,
  \dodoi{10.3847/2041-8205/819/2/L37}

\bibitem[{{Visvanathan} \& {Elliot}(1973)}]{1973ApJ...179..721V}
{Visvanathan}, N., \& {Elliot}, J.~L. 1973, ApJ, 179, 721,
  \dodoi{10.1086/151911}

\bibitem[{{Zdziarski} {et~al.}(1996){Zdziarski}, {Johnson}, \&
  {Magdziarz}}]{1996MNRAS.283..193Z}
{Zdziarski}, A.~A., {Johnson}, W.~N., \& {Magdziarz}, P. 1996, \mnras, 283, 193

\bibitem[{{Zu} {et~al.}(2011){Zu}, {Kochanek}, \&
  {Peterson}}]{2011ApJ...735...80Z}
{Zu}, Y., {Kochanek}, C.~S., \& {Peterson}, B.~M. 2011, \apj, 735, 80,
  \dodoi{10.1088/0004-637X/735/2/80}

\bibitem[{{{\.Z}ycki} {et~al.}(1999){{\.Z}ycki}, {Done}, \&
  {Smith}}]{1999MNRAS.309..561Z}
{{\.Z}ycki}, P.~T., {Done}, C., \& {Smith}, D.~A. 1999, \mnras, 309, 561,
  \dodoi{10.1046/j.1365-8711.1999.02885.x}

\end{thebibliography}

\end{document}